\documentclass[sn-basic]{sn-jnl}

\usepackage{natbib}
\usepackage{hyperref}
\usepackage{upgreek}
\usepackage{sidecap}
\usepackage[export]{adjustbox}  

\usepackage{siunitx}%
\DeclareSIUnit\ev{ev}
\DeclareSIUnit\px{pixel}

\usepackage{acronym}    
\acrodef{BL}[BL]{boundary layer}
\acrodef{CD}[CD]{contrast detection}
\acrodef{CFD}[CFD]{computational fluid dynamics}
\acrodef{muDH}[$\upmu$DH]{digital microscopic holography}
\acrodef{DNS}[DNS]{direct numerical simulation}
\acrodef{DFRH}[DFRH]{digital Fresnel reflection holography}
\acrodef{DVS}[DVS]{dynamic vision sensing}
\acrodef{DSR}[DSR]{dynamic spatial range}
\acrodef{DVR}[DVR]{dynamic velocity range}
\acrodef{EBIV}[EBIV]{event-based imaging velocimetry}
\acrodef{EBV}[EBV]{event-based vision}
\acrodef{FOV}[FOV]{field of view}
\acrodef{FWHM}[FWHM]{full-width-half-mean}
\acrodef{HFSB}[HFSB]{helium filled soap bubbles}
\acrodef{HIT}[HIT]{homogeneous and isotropic turbulence}
\acrodef{HP-LED}[HP-LED]{high-power LED}
\acrodef{HWA}[HWA]{hotwire anemometry}
\acrodef{IPCT}[IPCT]{image pattern correlation technique}
\acrodef{LED}[LED]{light emitting diode}
\acrodef{LDA}[LDA]{laser Doppler anemometry}
\acrodef{LES}[LES]{large-eddy simulation}
\acrodef{LPT}[LPT]{Lagrangian particle tracking}
\acrodef{MAmuPTV}[MA$\upmu$PTV]{multi-aperture micro-PTV}
\acrodef{PDF}[PDF]{probability density function}
\acrodef{PIV}[PIV]{particle image velocimetry}
\acrodef{ppp}[ppp]{particles per pixel}
\acrodef{PTV}[PTV]{particle tracking velocimetry}
\acrodef{PWM}[PWM]{pulse width modulation}
\acrodef{RMS}[rms]{root mean square}
\acrodef{2D-2C-PIV}[2D-2C-PIV]{two-dimensional (planar), two component particle image velocimetry}
\acrodef{ROI}[ROI]{region of interest}
\acrodef{STB}[STB]{"Shake the Box"}
\acrodef{TBL}[TBL]{turbulent boundary layer}
\acrodef{TCF}[TCF]{turbulent channel flow}
\acrodef{WSS}[WSS]{wall shear stress}

\jyear{2024}%

\begin{document}

\title[Dynamic wall shear stress by 3D-EB-PTV]{Dynamic Wall Shear Stress Measurement using Event-based 3D Particle Tracking}


\author{\fnm{Christian E.} \sur{Willert}}\email{chris.willert@dlr.de}
\author{\fnm{Joachim} \sur{Klinner}}\email{joachim.klinner@dlr.de}

\affil{\orgdiv{DLR Institute of Propulsion Technology}, \orgname{German Aerospace Center}, \orgaddress{\city{K\"{o}ln}, \postcode{51170}, \country{Germany}}}


\abstract{
    We describe the implementation of a 3d \ac{LPT} system based on \ac{EBV} and demonstrate its application for the near-wall characterization of a \ac{TBL} in air.
    The viscous sublayer of the \ac{TBL} is illuminated by a thin light sheet that grazes the surface of a thin glass window inserted into the wind tunnel wall. 
    The data simultaneously captured by three synchronized event-cameras is used to reconstruct the 3d particle tracks within \qty{400}{\um} of the wall on a field of view of \qtyproduct{12.0x7.5}{\mm}.
    The velocity and position of particles within the viscous sublayer permit the estimation of the local vector of the unsteady \ac{WSS} under the assumption of linearity between particle velocity and \ac{WSS}. 
    Thereby, time-evolving maps of the unsteady \ac{WSS} and higher order statistics are obtained that are in agreement with DNS data at matching Reynolds number. 
    Near-wall particle acceleration provide the rate of change of the \ac{WSS} which exhibits fully symmetric log-normal superstatistics.
    Two-point correlations of the randomly spaced \ac{WSS} data are obtained by a bin-averaging approach and reveal information on the spacing of near-wall streaks.
    The employed compact \ac{EBV} hardware coupled with suited \ac{LPT} tracking algorithms  provide data quality on par with currently used, considerably more expensive, high-speed framing cameras.
}




\maketitle


\section{Introduction}
\label{Sec:Intro}

\Acf{EBV}, also termed \emph{dynamic vision sensing} (DVS), is a new upcoming field within the field of computer vision and is inspired by the spiking mode of operation of the eye's retina.
Contrary to conventional frame-based imaging, \ac{EBV} only records changes of image intensity (i.e. contrast changes) on the pixel level, triggering a positive event ($+1$) for increasing intensity and a negative event ($-1$) for a decreasing intensity change. 
The typical threshold of the intensity-change trigger is on the order of 10--20\% but can be fine tuned.
As the pixels on the detector respond individually, the events appear asynchronously throughout the detector area resulting in a continuous stream of data, with each event datum $E_i = E_i(\mathbf{x},t,p)$ consisting of pixel coordinates $\mathbf{x_i} = (x_i,y_i)$, a time stamp $t_i$ and a polarity $p_i \in \{+1,-1\}$ indicating the direction of the intensity change. 
Unlike to conventional imaging, intensity is not directly available and the random nature of the asynchronous stream of events necessitates completely different data processing algorithms that are subject of current research.
For a recent review of the technology and underlying concepts the reader is referred to the topical review by \cite{EBVreview:2022}.

After original prototype and conceptual development of the technology in 1990's, affordable and ready-to-use hardware based on \ac{EBV} only recently have become available with current sensor resolutions of 1 MPixel.
This has broadened the range of applications as testified in a steadily increasing number of publications (see e.g. \citealp{GIT:EBVrefs,Gehrig:2024})).

The application of \ac{EBV} for the visualization and measurement of fluid flows is by no means new. 
Initial work was performed by \cite{Drazen:2011} on \ac{PTV} of dense particles in a solid–liquid two-phase pipe flow using an \ac{EBV} sensor of \numproduct{256x256} pixels and continuous laser (5W) illumination.
\cite{Ni:2012} used an \ac{EBV} array of 128$\times$128 elements to demonstrate micro-particle tracking ($\upmu$PTV) with \qty{12}{\um} microspheres and were able to detect Brownian motion.
Using a stereoscopic EBV system, \cite{Wang:2020} implemented a 3d PTV system allowing them to reconstruct three-dimensional tracks combining 2d tracking results from camera. Their flow experiment consisted of a small hexagonal cell with stirrer inducing a swirling flow containing O(\qty{100}{\um}) polystyrene spheres.
First PTV measurements in an air flow were performed by \cite{Borer:2017} using three synchronized EBV cameras (128$\times$128 pixels) to track \ac{HFSB} in volumes up to about 1\,m side length using white light LED arrays for illumination. The flow was only sparsely seeded allowing individual particles to be tracked with final data sets containing up to O(1\,000 - 10\,000) tracks.
More recently, \cite{Rusch:2023} re-implemented this concept as a real-time 3d \ac{PTV} system enabling live flow field reconstruction.

The work presented herein extends upon the recently introduced \ac{EBIV} concepts \citep{Willert_PulsedEBIV:2023,Willert_EBIV:2022} and introduces a 3d-3c \acf{LPT} system in a macroscopic imaging configuration with a magnification of O(10\,\unit{\um/pixel}), thereby capable of resolving the flow at the viscous scale. 
In comparison to previous event-imaging implementations, much higher seeding densities are achieved. 
However, due to the high data load, the captured sequences of event data currently cannot be processed in real-time and have to be analysed in an off-line fashion, that is, after completion of the measurement.

To demonstrate the viability of the proposed technique, the setup is used to acquire the near wall trajectories of tracers within the viscous sublayer of a \ac{TBL}, specifically to estimate the unsteady \ac{WSS}. 
In this sense the work addresses the current shortcoming of measurement techniques capable of providing reliable data of the unsteady \ac{WSS} vector.
The acquired data can be directly compared to readily available \ac{DNS} at matching Reynolds numbers for this canonical \ac{TBL} flow.

As pointed out in the review by \cite{OrluVinuesa:2020} the measurement of the unsteady \ac{WSS} remains a challenge with very few approaches capable of measuring the unsteady \ac{WSS} directly, with the exception of a few micro-mechanical implementations such as the shear stress imaging device by \cite{Kimura:MEMS:1999} which relies on the measurement of the actual shear force acting on an array of transducers.
The majority of \ac{WSS} measurement devices rely on an indirect measurement, typically of the near-wall velocity in the region dominated by viscous forces, namely the viscous layer at the wall, which extends out to about 5 viscous units $\mathit{l}^* = \nu\,/u_\tau$. Here, $\nu$ is the kinematic viscosity of the fluid and $u_\tau$ the friction velocity which itself is related to the \ac{WSS} $\tau_w$ and the fluid's density $\rho$ by $u_\tau = \sqrt{\tau_w / \rho}$.

Aside from hotwire anenometry (HWA) and the micro-pillar method \citep{Bruecker:MPS:2007,GrosseSchroeder:MPS3:2008}, most indirect \ac{WSS} measurement techniques are particle-based methods, that is, variants of \ac{LDA} or particle imaging. 
Among these, the following offer the desired combination of unsteady measurement of the \ac{WSS} vector on a reasonable \ac{FOV}, that is, they are not limited to point-wise measurements:
\begin{itemize}
    \item \cite{Gnanamanickam:2013}, \cite{LiuSchroeder:2019} and \cite{Bruecker:MPS:2015} used micro-pillars to get maps of the unsteady \ac{WSS}. Some of the measurements were biased because the length of the pillars extended beyond the viscous layer.
    \item Using \ac{muDH} \citep{Sheng:2008} or \ac{DFRH} \citep{Kumar:2021} the \ac{TBL} was imaged at a high magnification to retrieve unsteady 3d-3c flow data. Both techniques were applied in \ac{TCF} in water.
    \item Near-wall PIV at high magnification generally provide data on the streamwise \ac{WSS} component $\tau_{w,x}$ \citep{deSilva:2014,Willert:2018,Wang:WSS:2020}.
    \item Depth-from-defocus techniques, such as astigmatic $\upmu$PTV \citep{Fuchs:WSS:2023} or \ac{MAmuPTV} \citep{Klinner:2024} image the particle field with the optical axis aligned to the wall-normal direction $y$. 
    \item The \ac{STB} technique \citep{Schanz:2016} provides time-resolved 3d-3c \ac{LPT} data \citep{Schroeder:ISPIV:2015,Schroeder:Lisbon:2024} and, in terms of data provided, is most closely related to the approach presented herein.
\end{itemize}



The following article introduces a measurement configuration that can capture fields of the unsteady \ac{WSS} vectors by tracking the motion of particles within the viscous sublayer using the novel event-imaging approach. 
The paper is organized as follows: the specifics of the event-camera based 3d-3c system are followed by a description of the data processing including the employed particle tracking algorithm and an error assessment. 
The results section concentrates on the variety of data that can be derived from the particle tracking data including near-wall flow statistics and derived \ac{WSS}.
The discussion positions the herein introduced event-based 3d \ac{PTV} among existing approaches and addresses short-comings of particle-imaging-based \ac{WSS} estimation.

\section{3d-3c event-based tracking system}
\label{Sec:3dEBPTVsystem}
\begin{figure}[htb]
	(a)\hspace{0.45\columnwidth}(b)\\
	\centerline{ %
		\includegraphics[height=.32\textwidth]{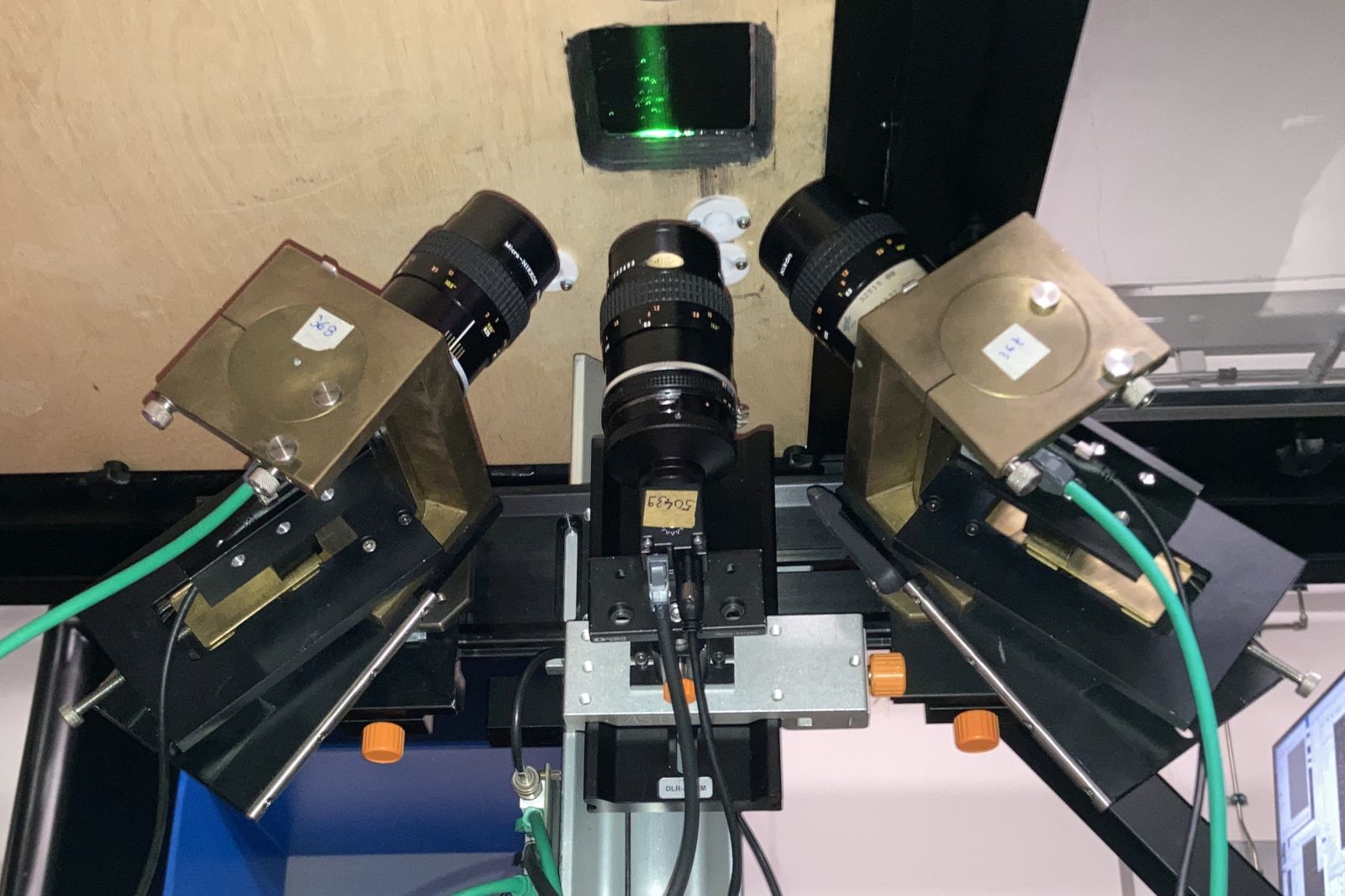}
		\hfill
		\includegraphics[height=.32\textwidth]{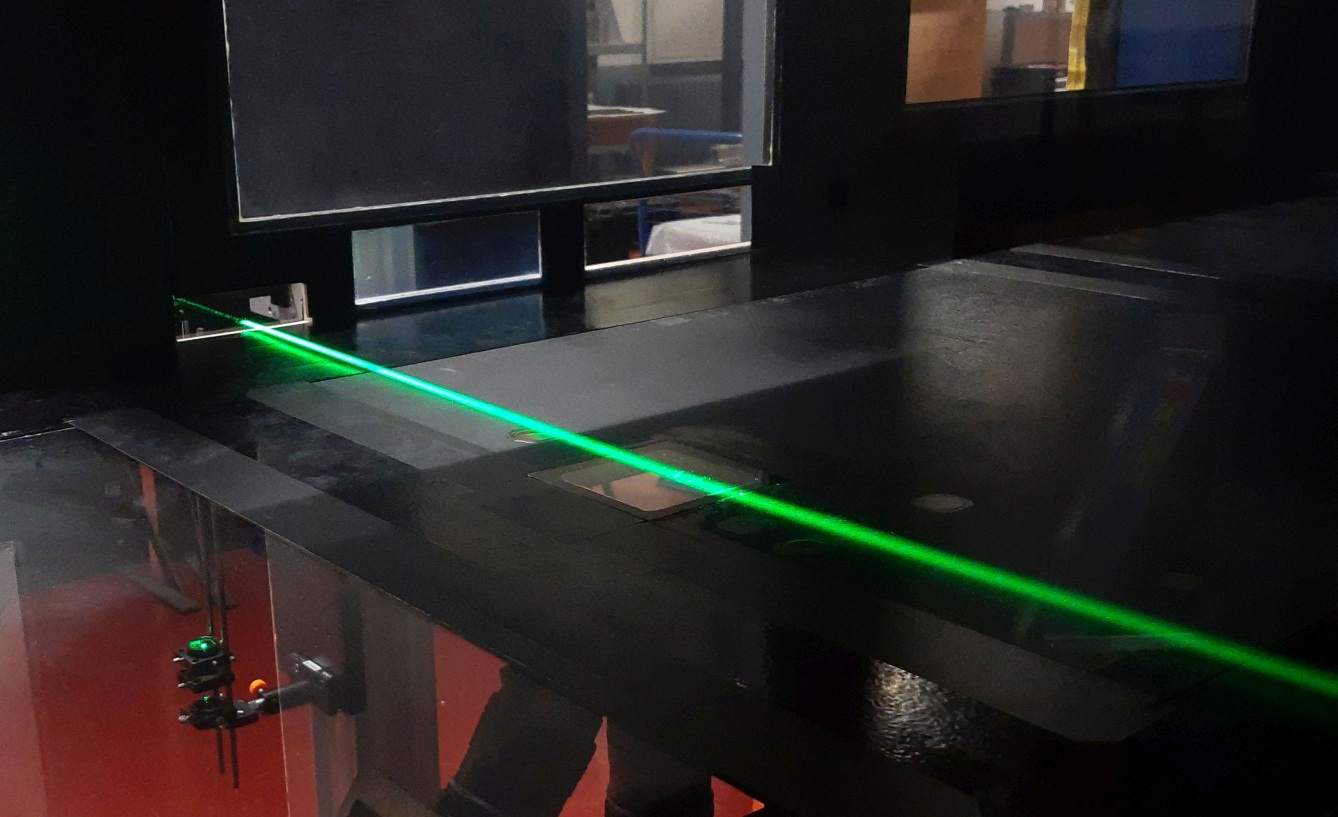}
	}
	\caption{Triple event-camera setup placed below the 1\,m wind tunnel of DLR in G\"{o}ttingen for particle tracking in the viscous sublayer of a \ac{TBL}, (b) laser light sheet grazing the window at the observation area at an estimated angle of $\approx$\qty{0.5}{\degree} to the surface.}\label{fig:EBIV_setup}
\end{figure}

\begin{SCfigure}[1][htb]
		\includegraphics[width=0.5\columnwidth]{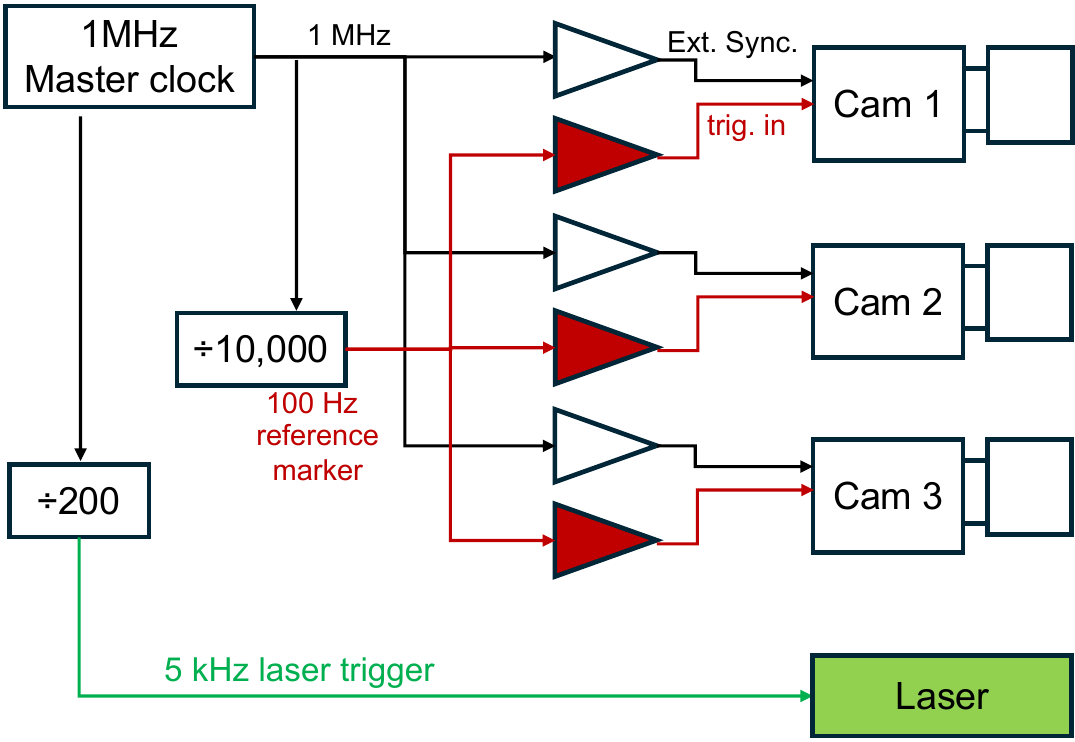}%
	\caption{Synchronization unit provides a common time base for all event cameras as well as a laser trigger.
	}
	\label{fig:camera_sync}
\end{SCfigure}

The 3d-3c particle tracking system comprises a triplet of event cameras (Prophesee EVK4, Sony IMX636 sensor, $1280 \times 720$ pixels, \qty{4.86}{\um} pixel size) in a photogrammetric configuration, that is, arranged in a manner to capture a common, relatively thin volume of interest. 
Scheimpflug mounts on the two off-normal cameras allow a common plane of focus for all three cameras (Fig.~\ref{fig:EBIV_setup}a).
The three cameras are synchronized with an external 1~MHz source to ensure common a time-base. 
In addition, reference pulses at 100\,Hz allow precise alignment of the separately recorded event sequences with a resolution of \qty{1}{\us} with respect to one another (cf. Fig.~\ref{fig:camera_sync}).
This is necessary since the cameras operate in a continuous mode and cannot be started from a common well-defined trigger and consequently require a posteriori re-alignment of the streamed data.

In the present application, the tracking system is mounted below the wind tunnel section and observes the bottom layer of the \ac{TBL} through a 1~mm thin glass window with anti-reflective coating. 
This domain is illuminated with a $\approx 0.5$~mm thin light sheet introduced from the side of the wind tunnel with a slight inclination ($\approx 0.5^\circ$, cf. Fig.~\ref{fig:EBIV_setup}b).
The light sheet is oriented such that all cameras receive the light scattered by the tracers at a common scattering angle of $90^\circ$. This results in similar illumination intensities on all three detectors and avoids angle-dependent Mie scattering differences between the cameras.

At a working distance of about 200 mm a common \ac{FOV} of about $12.0 \times 7.5 \,\textrm{mm}^2$ is captured (magnification $m = 0.48$ with 10\,$\upmu\textrm{m/pixel}$).
The pulsed laser (Innolas Nanio-Air) is operated at pulsing frequency of  5~kHz with an integral power of about 1--2\,W and is synchronized to the camera time base (see Fig.~\ref{fig:camera_sync}).
The macro objective lenses (Nikon Micro-Nikkor 55 mm / 2.8) are stepped down to $f_\# = 8$.

Water-based tracer particles of about $1-2\,\upmu\textrm{m}$ and a life-time of about 10 minutes are provided by a fog generator (HazeBase Classic, base*M fluid).
Event-recordings of up to 60\,s duration are acquired at wind tunnel speeds of $U_\infty = \qty{5.2},\,\qty{7.5}\, \textrm{and} \,\qty{10} \,\unit{\m\per\s}$. 
Table~\ref{tbl:event_data} summarizes specific aspects of the acquired raw data such as event data rate and the amount of actual data streamed to the host computer. 
Table~\ref{tbl:event_data} also provides the \ac{TBL}'s characteristic parameters which were obtained by high-speed profile PIV \citep{Klinner:2024}.

\begin{figure}[htb]
	 (a)\hspace{0.46\columnwidth}(b)\\[4pt]
	\centerline{ %
		\includegraphics[width=.48\textwidth,frame]{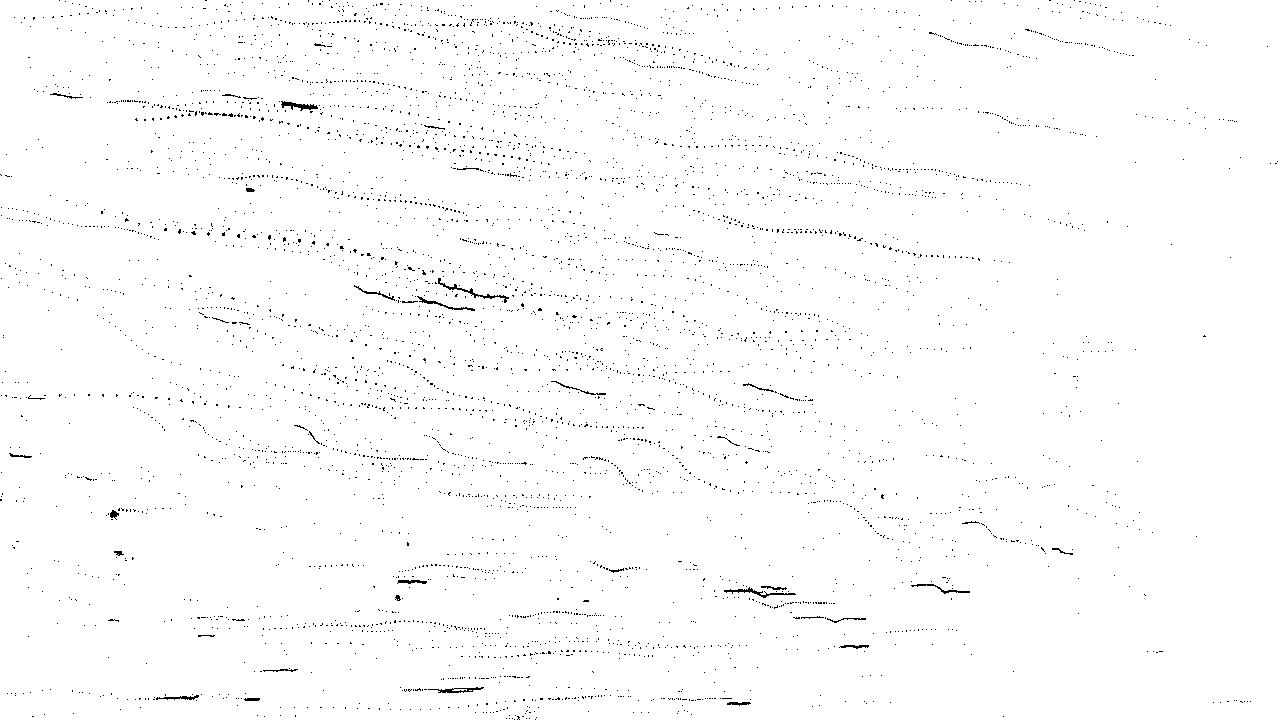}
		\hfill
		\includegraphics[width=.48\textwidth,frame]{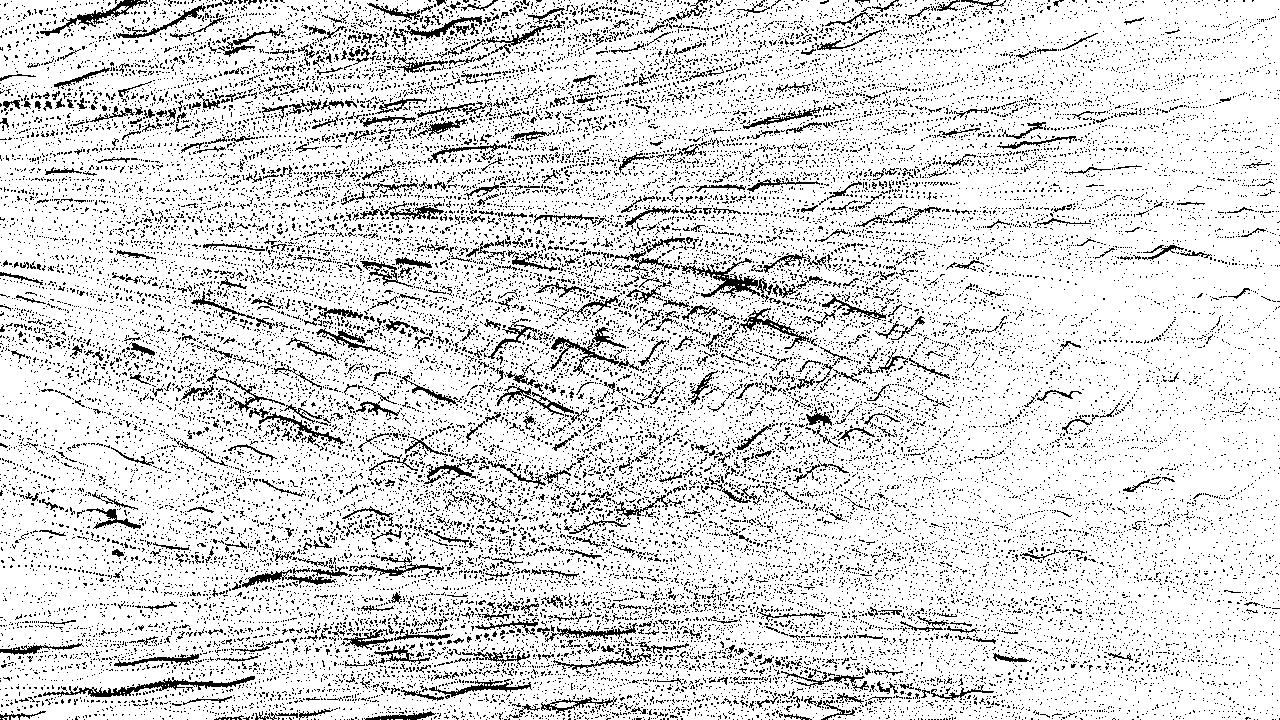}
	}
	 (c)\hspace{0.46\columnwidth}(d)\\[4pt]
 \centerline{ 
     \includegraphics[trim=220 200 460 200,clip, width=0.48\textwidth,frame]{./Fig3a_nearwall5mps3_cam00050439_2023-02-02_18-33-31_731ms_neg}
     \includegraphics[trim=220 200 460 200,clip, width=0.48\textwidth,frame]{./Fig3b_nearwall5mps2_cam00050439_2023-02-02_18-16-31_146ms_neg}
}
	\caption{Sample pseudo-images from \qty{10}{\ms} of the event stream recorded by the central camera at \qty{1.5}{\mega\ev/\s} (a) and \qty{23}{\mega\ev/\s} (b). 
 The lower row (c) and (d) show zoomed portions of the above. With a laser pulse rate of \qty{5}{\kHz}, the pseudo-images contain 50 laser pulses.}\label{fig:EBIV_rawdata}
\end{figure}

Keeping the laser energy and light sheet position constant, the seeding density is varied by more than one order of magnitude resulting in a corresponding variation of the event data rates.
While the camera settings (so called \textit{biases}) have been tuned to favor positive ($+1$ events), the percentage of negative ($-1$ events) shows a gradual increase with increasing event-rate. This is a specific behavior of the event-detector itself, a further investigation thereof being beyond the scope of this article. 
Fig.~\ref{fig:EBIV_rawdata} intends to provide an impression of the event data acquired by one of the camera at two different seeding concentrations.

\begin{table}[tb]
    \caption{Overview of acquired event data including characteristics of the studied \ac{TBL} as determined with HS profile PIV.
    }
    \label{tbl:event_data}
    \centering
    \begin{tabular}{ll cccc|ccc|c}
    \toprule
    $\mathbf{U}_\infty$ & $[\unit{m\per\s}]$ & \multicolumn{4}{c|}{\textbf{5.2}} & 
\multicolumn{3}{c|}{\textbf{7.5}} & \textbf{10.0} \\[3pt]
      Data set & & 5-3 & 5-1 & 5-4 & 5-2 & 7-1 & 7-3 & 7-2 & 10-1 \\
    \midrule
    Duration & [s] &
        60 & 60 & 60 & 60 & 10 & 10 & 10 & 10 \\ 
        Event rate & [$10^6$ \unit{Ev\per\s}] & 1.5 & 8.2 & 14.1 & 23.2 & 8.1 & 16.6 & 16.9 & 13.4 \\
    Pos. Events & & 97\,\% & 85\,\% & 77\,\% & 75\,\% & 89\,\% & 80\,\% & 80\,\% & 81\,\% \\
    Data rate$^1$ & [\unit{MB\per\s}] & 21 & 92 & 148 & 216 & 92 & 167 & 169 & 141 \\
    Track yield$^2$ & & 60\,\% & 54\,\% & 39\,\% & 4.8\,\% & 
    40\,\% & 30\,\% & 24\,\% & 41\,\% \\
    \midrule
    $\mathbf{Re}_\tau$ & & \multicolumn{4}{c|}{\textbf{563}} & \multicolumn{3}{c|}{\textbf{754}} & \textbf{935} \\
    $\mathit{u}_\tau$ & $[\unit{m\per\s}]$ & \multicolumn{4}{c|}{0.223} & \multicolumn{3}{c|}{0.304} & {0.390} \\
    $\mathit{l}^* = \nu\,/u_\tau$ & $[\unit{\um}]$ & \multicolumn{4}{c|}{68.8} & \multicolumn{3}{c|}{50.5} & {39.4} \\
    $\delta_{99}$ & $[\unit{\mm}]$ & \multicolumn{4}{c|}{38.7} & \multicolumn{3}{c|}{38.1} & {36.8} \\
      \bottomrule
    \end{tabular}
    \\ $^1)$ combined for all three event-cameras,
    \\ $^2)$ tracks of length $N_\mathrm{track} \geq 7$ after validation at a wall distance $[0.5 < y^+ 1.5]$.
\end{table}

\subsection{Tracer characteristics in the viscous layer}
\label{sec:tracers}
As pointed out by \cite{ShihLumley_Kolmogorov:1993} the Kolmogorov velocity scale $u_\eta$ near the wall is of same magnitude as the friction velocity $u_\tau$ such that the length scale $\eta$ is given by
\begin{equation}
    \label{eq:KolmogorovVelocity}
    \frac{\nu}{u_\tau} \approx \frac{\nu}{u_\eta} 
    = \frac{\nu}{(\eta\epsilon)^{1/4}} = \eta .
\end{equation}
Under the same premise, the Kolmogorov time scale $t_\eta = (\nu/\epsilon)^{1/2}$ is related to the shear strain rate at the wall 
\begin{equation}
    \label{eq:KolmogorovTime}
    \left.\dfrac{\partial {u}}{\partial y}\right\rvert_{y=0} 
    = \dfrac{{u_\tau}^2}{\nu} 
    \approx \frac{{u_\eta}^2}{\nu} = \frac{(\nu\epsilon)^{1/2}}{\nu}
    = \frac{1}{t_\eta} .
\end{equation}
With the Kolmogorov microscales directly corresponding to the viscous scales the following quantities are obtained for the $U_\infty = 5.2\,\unit{\m\per\s}$ flow condition:
$\eta \approx 70\,\unit{\um}$, $t_\eta \approx 300\,\unit{\us}$ and $u_\eta \approx 0.22\,\unit{\m\per\s}$.
These quantities are of relevance for characterization of tracer particle performance described next.

A quantity describing a particle's fidelity of moving with the flow, that is, along the streamline, is the Stokes number given by the ratio of the particle response time $t_p$ and the characteristic time scale of the flow $t_f$:
\begin{equation}
    \label{eq:StokesNumber}
    \mathrm{Stk} = \frac{t_p}{t_f} 
    \;\; \mathrm{with} \;\;
    {t_p} = \frac{1}{18}\frac{\rho_p d_p^2}{\mu_f}
\end{equation}
for spherical particles of diameter $d_p$ and density $\rho_p$ carried in a fluid with dynamic viscosity $\mu_f$.
Values significantly smaller than $\mathrm{Stk} = 1$ indicate a good flow tracking performance.

The water-glycol droplets used in the present 3d \ac{LPT} measurements have a size range of $[1\,\unit{\um} < d_p < 2\,\unit{\um}]$ with corresponding relaxation times of  $[3\,\unit{\us} < t_p < 12\,\unit{\us}]$. 
In the viscous sublayer $t_f = t_\eta$ such that the Stokes number becomes $\mathrm{Stk} \leq 0.04$ at $U_\infty = 5.2\,\unit{\m\per\s}$. 
At $U_\infty = 7.5\,\unit{\m\per\s}$ the characteristic time scale decreases to $t_\eta \approx 165\,\unit{\us}$ with $\mathrm{Stk} \leq 0.07$. 
At the highest velocity of $U_\infty = 10.0\,\unit{\m\per\s}$ and $t_\eta \approx 100\,\unit{\us}$ the Stokes number further increases to $\mathrm{Stk} \leq 0.12$ for particles with $d_p \leq 2\,\unit{\um}$.
Overall this indicates an adequate tracking performance, especially at the lower tunnel operating condition.

\section{Camera calibration}
\label{sec:camcalib}
Elemental for reliable 3d-PTV is an accurate camera mapping which allows a transformation from image space into object space and back. This is generally achieved using established camera calibration procedures. 

\begin{SCfigure}[1][htb]
	(a)  \includegraphics[height=.4\textwidth]{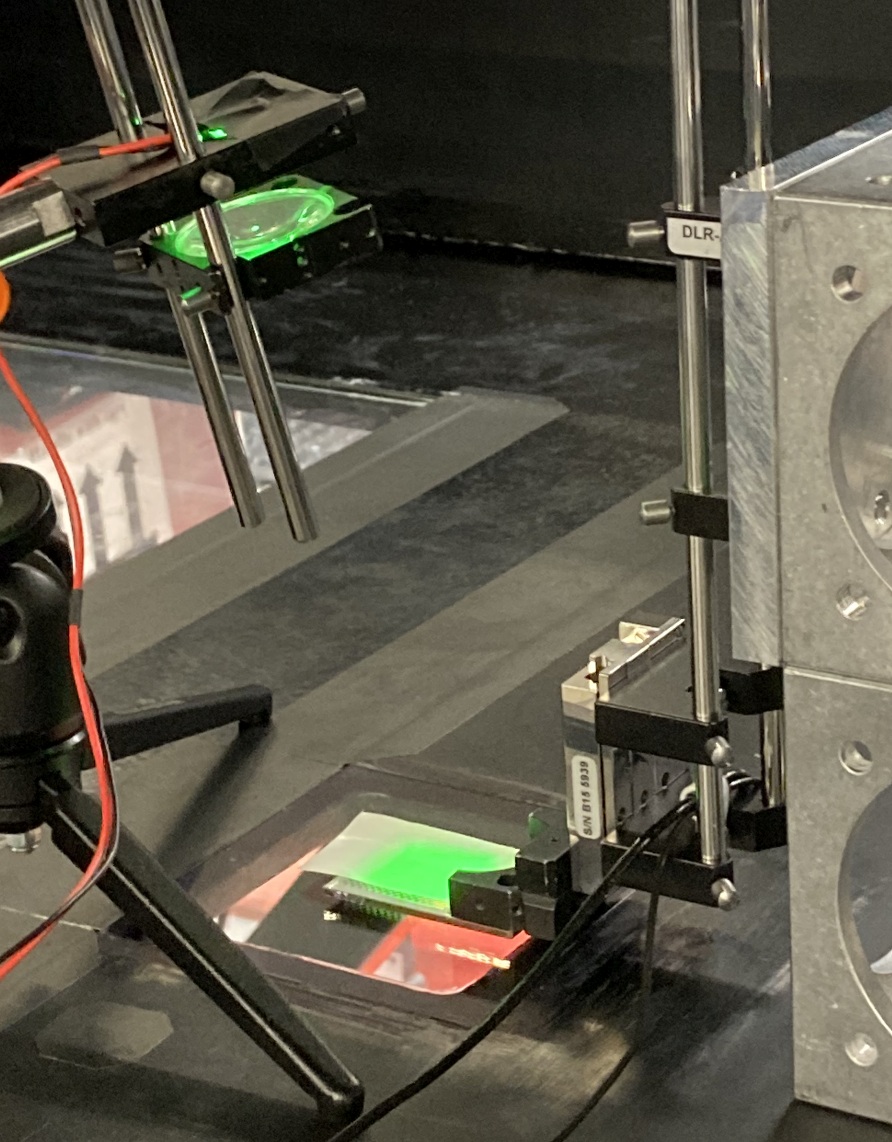}
	\hspace{2mm}
	(b)  \includegraphics[height=.4\textwidth]{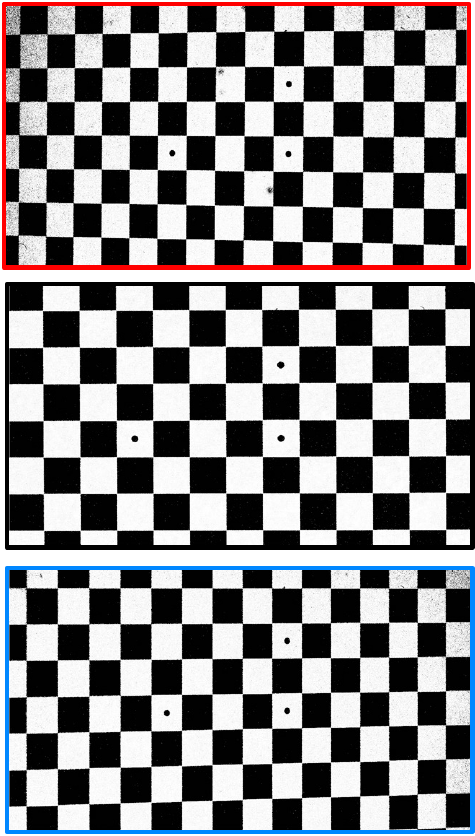}
	\caption{Camera calibration setup using a back-illuminated checker-board target mounted on a micro-translation stage (a), simultaneous camera views of the target (b).}\label{fig:EBIV_calib}
\end{SCfigure}

\begin{figure}[htb] 
	\centering
	(a)  \includegraphics[width=0.75\textwidth]{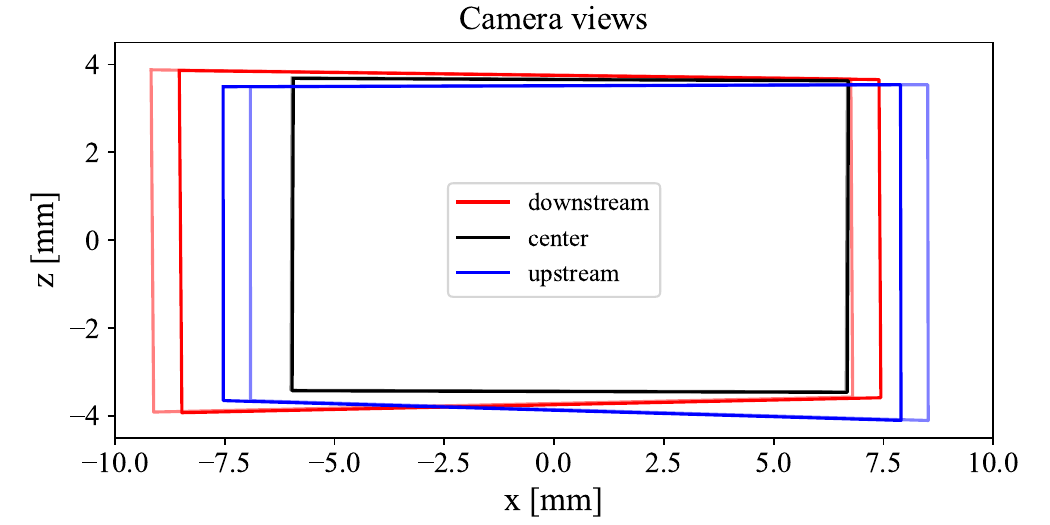}\\
    (b)  \includegraphics[trim=30 40 30 40,clip, width=0.75\textwidth]{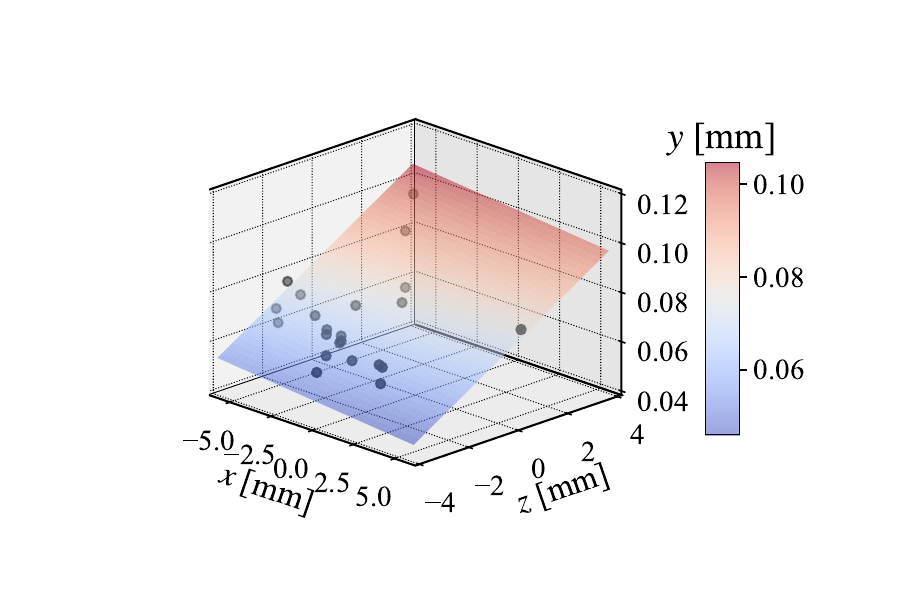}
	\caption{Camera field of view at $y=0$ and $y=500\,\unit{\um}$ (a).
    Reconstructed plane of the glass insert based on 3d reconstruction of stationary particles stuck to the surface (b).}\label{fig:cam_views}
\end{figure}

Calibration data in the form of image-object correspondence points is collected from recordings of a calibration target. 
Here, a checker-board target with \qtyproduct{1x1}{\mm} squares printed on glass is mounted parallel to the observation window and traversed in wall-normal ($y$) direction at increments of $\Delta y = 250\,\unit{\um}$ (Fig.~\ref{fig:EBIV_calib}a). 
Due to insensitivity of the \ac{EBV} to static imagery, the glass target is back-illuminated by a pulsed LED at \qty{100}{\Hz}. 
Summing events over a period of \qty{0.5}{\s} provides high-contrast calibration images suitable for grid marker detection (Fig.~\ref{fig:EBIV_calib}b).
The common \ac{FOV} shared by the cameras, as depicted in Fig.~\ref{fig:cam_views}, extends about \qty{12}{\mm} by \qty{7.5}{\mm} in streamwise and spanwise direction, respectively.

The accuracy of particle reconstruction in relation to the glass surface requires the knowledge of the plane spanned by the target in relation to the plane of glass surface.
This is achieved by triangulation of stationary particles and dust attached to the surface of the window, which are readily detected in the raw event data as continuous triggered pixel clusters. 
A 2d plane fit provides the reference plane to which the reconstructed track data will be aligned (Fig.~\ref{fig:cam_views}b). 
The slope amounts to about \qty{50}{\um} across the \qty{10}{\mm} \ac{FOV} (inclination $\approx$\qty{0.3}{\degree}).

A dual plane method is used to map between object and image space and to compute epipolar lines to match the particle images between the views. 
A particle-based residual alignment such as typically performed in 3d \ac{STB} \ac{LPT} is currently not applied.

\section{Event-data processing}
\label{sec:processing}
\begin{figure}[tb]
	{ \small{(a)}\hspace{0.49\textwidth}{(b)}}\\
	\centerline{ %
		\includegraphics[width=.49\textwidth]{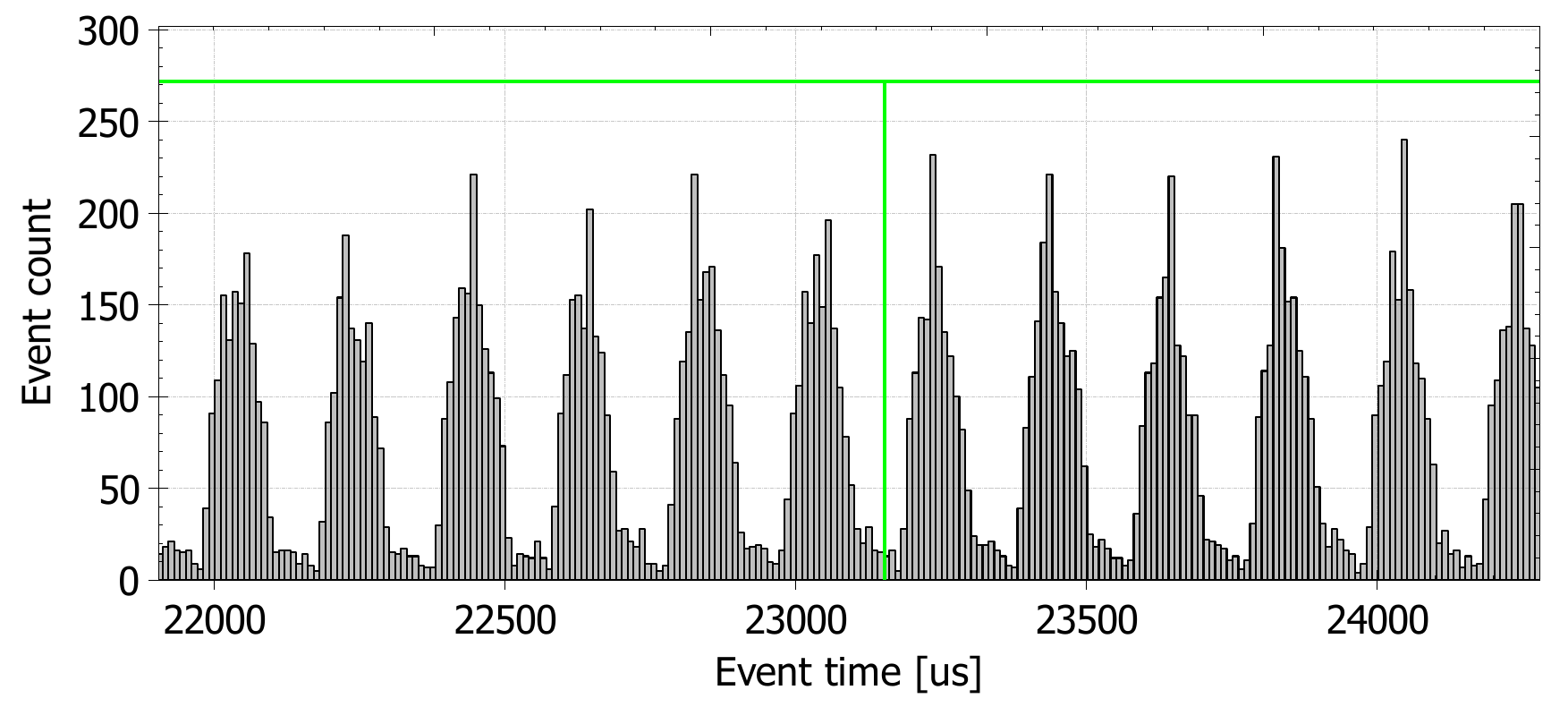}
		\hfill
		\includegraphics[width=.49\textwidth]{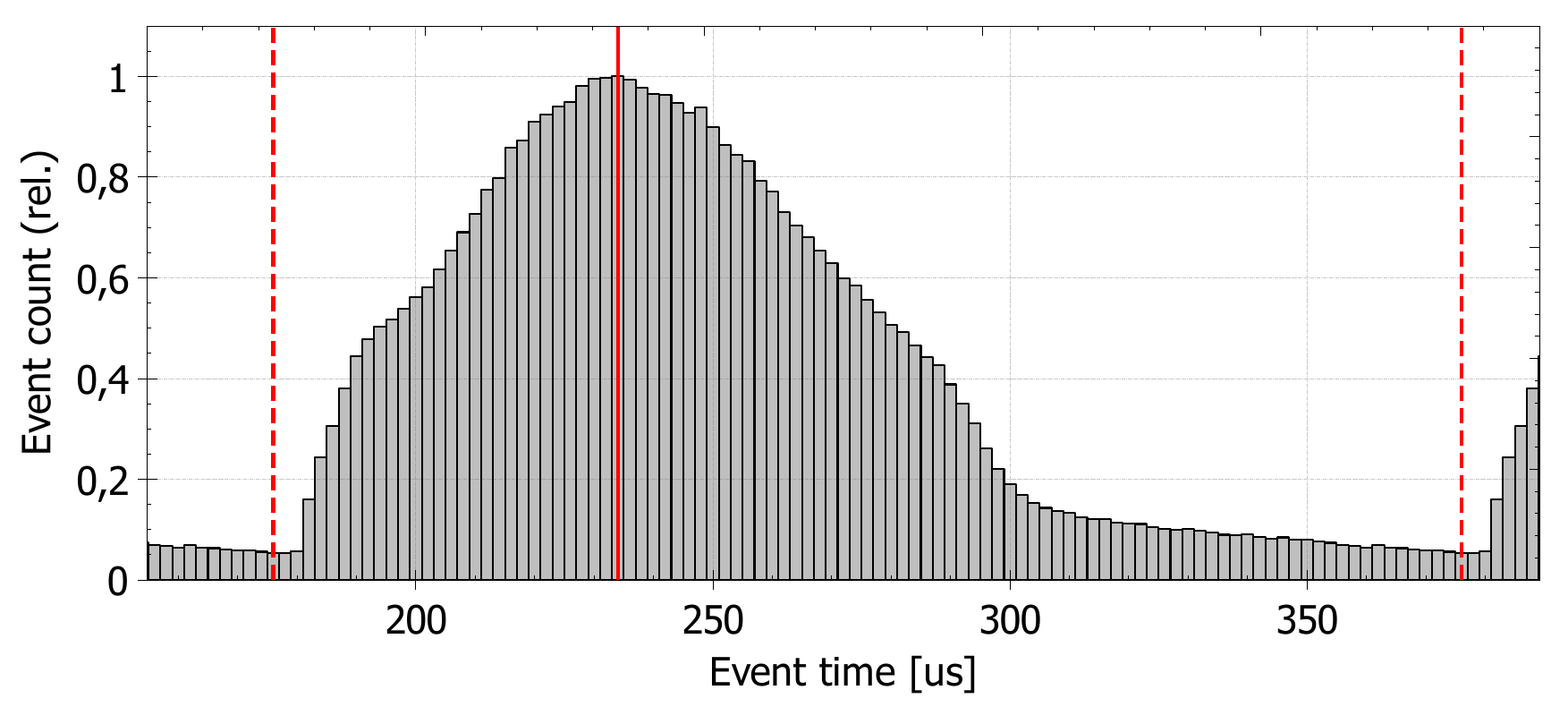}
	}
	\caption{Histograms of positive ($+1$) event data recorded by one of the three cameras with the laser pulsing at \qty{5}{\kHz}; green line indicates a periodic reference marker at \qty{100}{\Hz} used for registration of event streams to one another; {(a)} raw stream binned at \qty{10}{\us} intervals, {(b)} mean event distribution during one pulsing period (\qty{200}{\us}).}\label{fig:raw_event_data}
\end{figure}

Prior to particle tracking, the acquired event-recordings are temporally aligned using the external 100\,Hz reference markers and then converted to pseudo-image sequences by re-sampling the event-data at a frequency corresponding to the laser pulsing rate. 
During a sample interval, e.g. \qty{200}{\us} at \qty{5}{\kHz} laser pulsing, any given pixel is only allowed to produce at most one event. Hence, the resulting pseudo-image is binary in nature. 
The automated event-sampling is performed on the basis of searching for the minimum in the ensemble averaged event histogram, a representative example of which is given in Fig.~\ref{fig:raw_event_data} for a laser pulsing frequency of \qty{5}{\kHz}. In this case, the sampling period, as indicated by the red dashed lines, would begin with an offset of $\approx$ \qty{175}{\us} and end \qty{200}{\us} later.
As described in more detail in \cite{Willert_PulsedEBIV:2023}, the use of pulsed light intends to mitigate issues related to the delayed response, e.g. latency, of the event detector. This latency is apparent in Fig.~\ref{fig:raw_event_data}(b) with events being registered by the detector up to \qty{100}{\us} after being exposed to the short pulse of light (about \qty{20}{\ns}). Another advantage of the pulsed light approach is the capture of stationary particles which would be ``invisible" when illuminated by a continuous light source. 
Finally, the use of continuous illumination was found to ``favor" slower particles since they have a higher likelihood of triggering events while crossing a given pixel. 
This would bias the measurement toward lower velocities already at the raw data stage \citep{Willert_PulsedEBIV:2023}.

Particle tracking is performed for each camera view individually by first extracting contiguous binary pixel blobs from the pseudo images and computing their centroids (center of mass). 
A k-d tree based nearest-neighbor search scheme then detects tracklets across three adjacent pseudo-images and extends these via a predictor scheme to the following image frames.
The tracker accepts gaps of up to one pseudo-frame to prevent a premature truncation of tracks. 
Using the 2d tracks for the three cameras views, reconstruction of the 3d tracks is performed using the epipolar lines of a given particle on the other two views.
Finally, a least-squares regression is applied to the reconstructed tracks to estimate a given particle's 3d position, velocity and acceleration along its path.

Using a cubic B-spline fit on the raw 3d position data yield a set of spline coefficients and provide a continuous description of the particle's position, velocity and acceleration along the the track, that is, in time and space.
The spline fit is weighted proportionally to the inverse of the residuals of the 3d particle position.
Track validation is based on the residuals of the cubic spline fit ($r_\mathrm{fit} \leq \qty{10}{\um}$), a minimal track length ($N_\mathrm{track} \geq 7)$ and maximum allowed wall-normal fluctuation ($v_\mathrm{rms} < 0.02 U_\infty$ for $y^+ < 2$).

The wall shear stress vector $\vec{\tau}_w = [\tau_{w,x},\tau_{w,z}]$ for each validated particle position is then obtained be dividing its estimated wall-parallel velocity $\mathbf{u} = [u,0,w]$ by its distance from the wall $\Delta y$ as an approximation to the definition of \ac{WSS} 
\begin{equation}
\label{eq:shearstress}
	\vec{\tau}_w = \mu \left.\dfrac{\partial \mathbf{u}}{\partial y}\right\rvert_{y=0} 
        =  \mu \lim_{y \rightarrow 0} \frac{ \mathbf{u}(y)}{y}
	\approx \mu\, \dfrac{\mathbf{u}(\Delta y)}{\Delta y}    
\end{equation}
with $\mu$ representing the dynamic viscosity and the range of $\Delta y$ limited the viscous layer ($y^+ < 5$).
Here it is crucial that the wall distance $\Delta y$ is corrected for any offset and tilt of the wall surface as described in Sec.~\ref{sec:camcalib}.

\subsection{Error estimation}
\label{sec:error_estimation}

The estimation of the wall shear stress $\tau_w$ based on the discrete approximation given in Eq.~\ref{eq:shearstress} is affected by two primary sources of error: (1) the uncertainty of the distance of the particle from the wall $\epsilon_y$ and (2) the measurement uncertainty in the particle's velocity $\epsilon_u$.
In combination, the two errors will result in a rapid increase of the measurement uncertainty toward the wall.
Classical methods of error propagation provide the relative error of the estimated \ac{WSS} for the approximation given in Eq.~\ref{eq:shearstress}:
\begin{eqnarray}
    \epsilon_{\tau_w}(y) = \frac{\delta\tau_w}{\tau_w} 
     & = &\left[{\epsilon_u}^2 + {\epsilon_y}^2\right]^{0.5} \\
     & = & \left[ \left(\frac{ \delta u(y)}{u(y)}\right)^2 + \left(\frac{ \delta y}{y}\right)^2 \right]^{0.5}
\end{eqnarray}

\begin{SCfigure}[1][htb]
	\includegraphics[width=.5\textwidth]{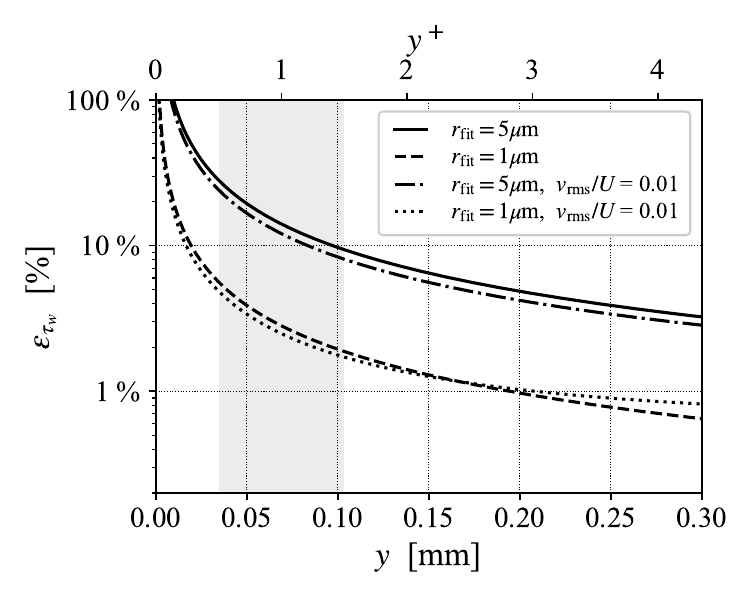}
	\caption{Estimated error of the \ac{WSS} for different mean values of the residuals of the 3d cubic B-spline track fitting.}\label{fig:error_estimate}
\end{SCfigure}

Reasonable constraints for $\delta y$ and $\delta u(y)$ can be derived from the residuals of the three-component reconstruction or the track-fitting scheme, here, a cubic B-spline. 
After validation, the latter are in the order of $r_\mathrm{fit} = \qty{5}{\um}$ in the sampling domain $[0.5 < y^+ < 1.5]$ (see Fig.~\ref{fig:velocity-profile}d).
As depicted in Fig~\ref{fig:error_estimate}, the error on a single \ac{WSS} estimate can be in excess of \qty{10}{\percent} in the domain of interest.
These estimates are based on the mean residual of the B-spline curve fit, $r_\mathrm{fit}$, obtained by subtracting the 3d track particle positions from their fitted positions.
Very close to the wall ($y^+ < 2$) the particle motion has a negligible wall-normal component $v$ such that the error can be constrained in the validation step by limiting the variance of $v$ (dash-dot and dotted curves in Fig~\ref{fig:error_estimate}). However, this has only a minor influence on the \ac{WSS} error $\epsilon_{\tau_w}$.

Using the single sample uncertainty $\sigma_{\tau_w} = \epsilon_{\tau_w} \tau_w$, the uncertainty on the mean and higher order statistics of a quantity $x$ can be expressed in terms of the \qty{95}{\percent} confidence interval \citep{Benedict:1996}:
\begin{eqnarray}
    \label{eq:confidence_mean}
    \epsilon_{\langle x \rangle} & = & \frac{1.96}{\langle x \rangle} 
    \sqrt{\frac{\sigma_x^2}{N_{s,\mathit{eff}}}}\\
    \label{eq:confidence_var}
    \epsilon_{x_\mathrm{rms}} & = & \frac{1.96}{{\langle x x \rangle}^{0.5}} 
    \sqrt{\frac{\sigma_x^2}{2 N_{s,\mathit{eff}}}}
\end{eqnarray}
where $N_{s,\mathit{eff}} < N_s$ is the estimated number of uncorrelated samples within the set $N_s$ and can be obtained by accounting for the integral time scale $T_u$ for the near-wall flow.
Using DNS, \cite{QuadrioLuchini:2003} estimated the integral time sale $T_u$ in the viscous layer at about 20 viscous time scales, $\mathit t^* = \nu\,u_\tau^{-2} \approx \tau_\eta$ (see Eq.~\ref{eq:KolmogorovTime}), such that $T_u \approx 20\, t^* \approx \qty{6}{ms}$ at $U_\infty = \qty{5.2}{\m\per\s}$.
This time-span covers 30 laser pulses (at $f_s = \qty{5}{\kHz}$) such that the effective number of samples becomes:
\begin{equation}
\label{eq:N_eff}
    N_{s,\mathit{eff}} = \frac{1}{T_u f_s} N_s =  \dfrac{1}{30} N_s .
\end{equation}
For $U_\infty = \qty{7.5}{\m\per\s}$ and $U_\infty = \qty{10}{\m\per\s}$, the factor ${T_u f_s}$ respectively reduces to 16.4 and 10.0.
Combining Eqs.~\ref{eq:confidence_mean} and \ref{eq:confidence_var} with Eq.~\ref{eq:N_eff} yields the uncertainty estimates for the given quantities. 
Even with a single sample uncertainty of $\epsilon_x = \qty{20}{\percent}$, the large sample sizes of O(\qty{1E7}) reduce the relative uncertainty to levels of O(\qty{1E-3}) for the \ac{RMS} values. 
Here it should be noted that this uncertainty does not include more influential bias errors that are e.g. introduced by possible misalignment of the estimated wall plane as well as vibrations and other calibration related errors.

\begin{figure}[tb]
	\centering
	\includegraphics[width=\columnwidth]{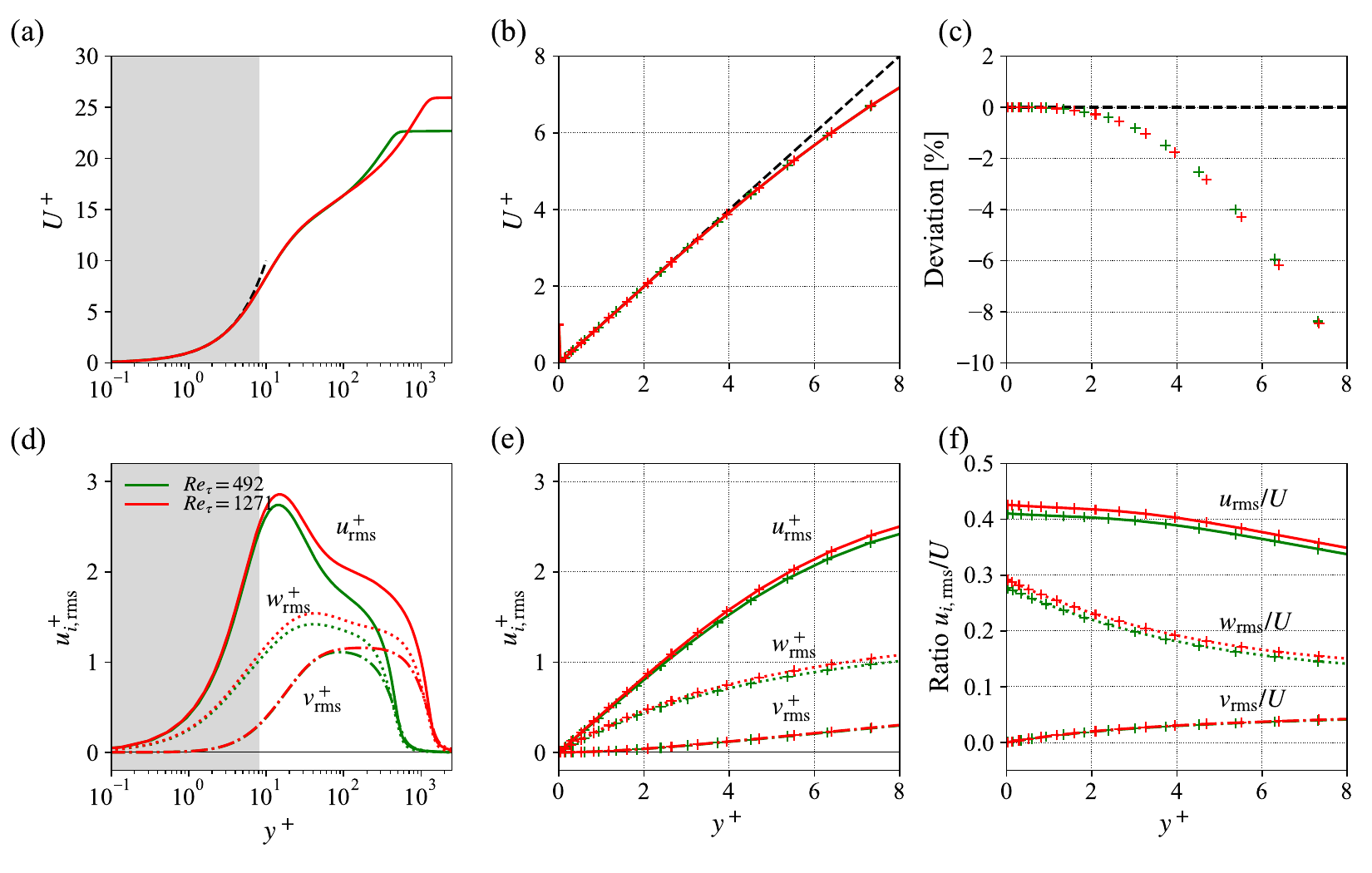}
	\caption{Profiles of mean streamwise $U^+$ (top row) and \acf{RMS} of all three velocity components (bottom row) from DNS of \ac{TBL}s by \protect\cite{Sillero:2013} in log-scaling (a,d), linear scaling near the wall (b,d). (c): deviation of $U^+$ from linearity; (f): velocity \acf{RMS} normalized with $U^+$. Gray shaded areas in (a,d) represent domains in (b,c,e,f) for $y^+ \leq 8$.}
	\label{fig:DNS_nearwall_rms}
\end{figure}

Another source of uncertainty is routed in the linearity assumption of the fluid velocity within the viscous sublayer. As illustrated in Fig.~\ref{fig:DNS_nearwall_rms}(c) for \ac{DNS} \ac{TBL}, 
the profile of mean streamwise velocity already deviates by nearly \qty{4}{\percent} from linearity at a wall distance of $y^+ = 5$. 
Therefore, a reliable estimation relies on particle velocity data provided for $y^+ \leq 4$.
At the same time, the relative uncertainty of $y_p$ rapidly increases as the particle distance $\Delta y$ approaches the wall as described before. 

The availability of DNS also provides justification for track validation based on the variances of the individual velocity components along the track. 
In particular, Fig.~\ref{fig:DNS_nearwall_rms}(e) shows that at a wall distance of $y^+ < 2$, the wall-normal fluctuations $v_\mathit{rms}$ are less an order of magnitude smaller than streamwise or spanwise fluctuations.
More importantly the velocity fluctuations, shown in Fig.~\ref{fig:DNS_nearwall_rms}(f), converge differently toward their limiting values at the wall, which has a notable influence on the estimation of the \ac{WSS} fluctuations as described later.

\section{Results}
\label{sec:results}
\begin{figure}[htb]
	\small
	{(a)}\hspace{.49\textwidth}{(b)}\\
	\centerline{
		\includegraphics[width=.49\textwidth]{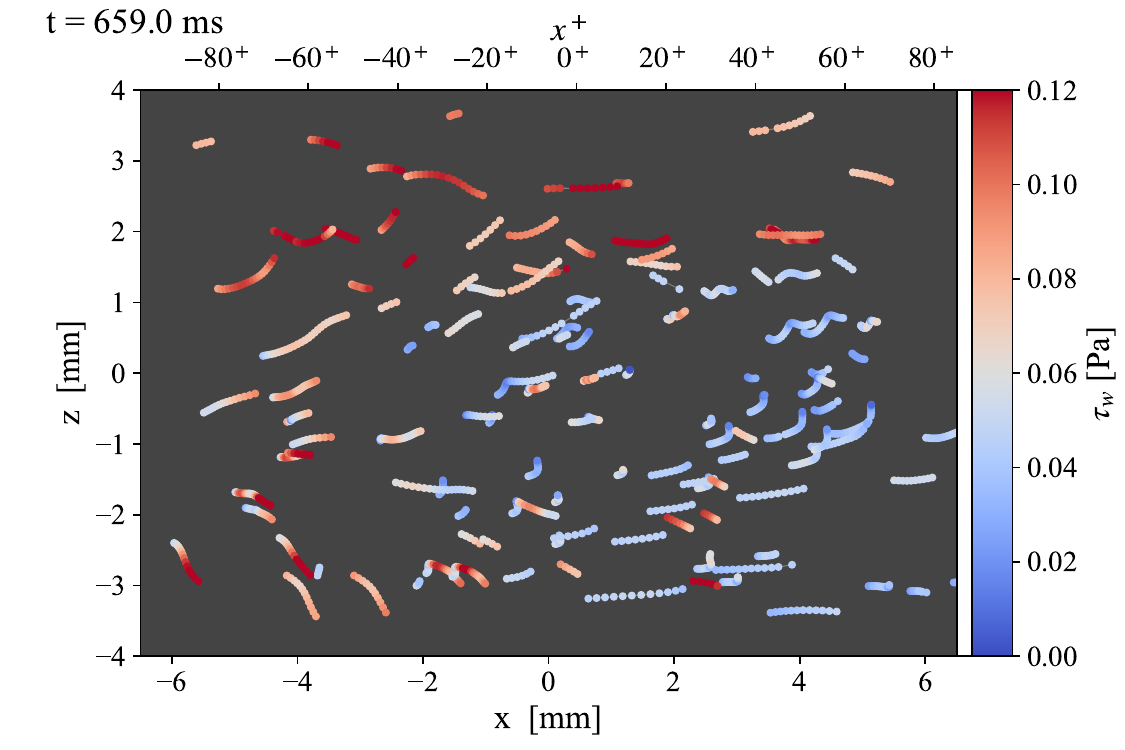}
		\includegraphics[width=.49\textwidth]{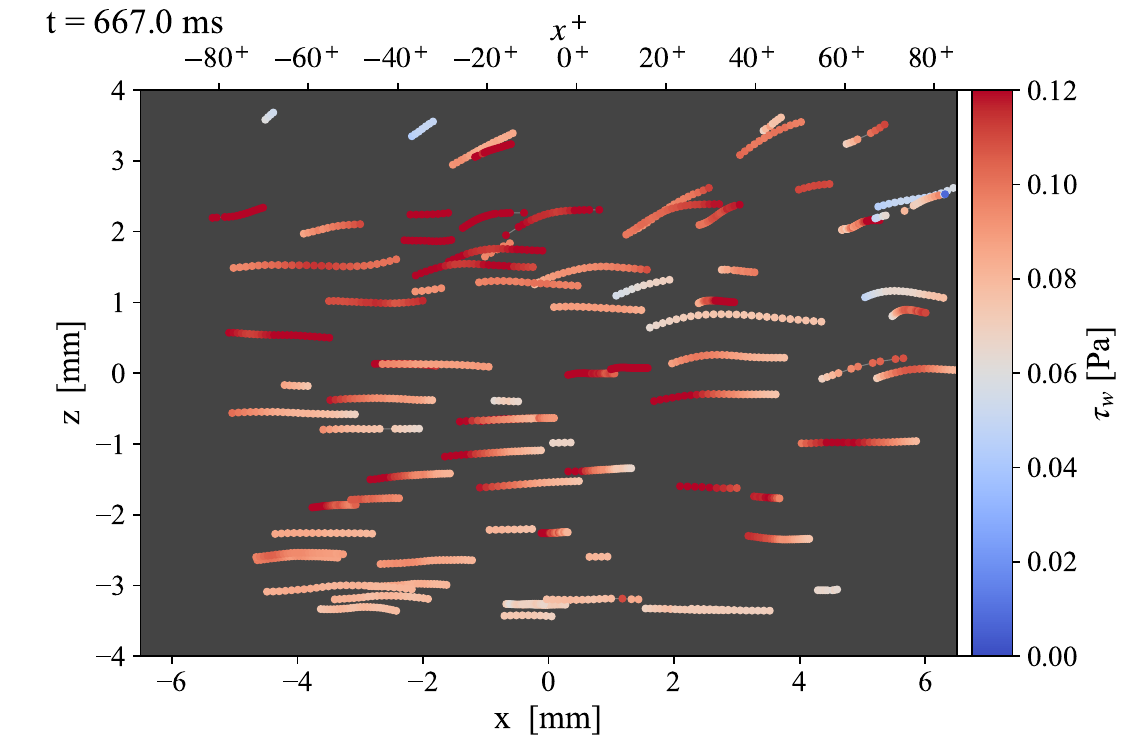}}
	\caption{Processed particle tracks color coded with wall shear stress magnitude at $\mathrm{Re}_\tau = 563$ ($U_\infty = \qty{5.2}{\m\per\s})$. Each frame represents 5\,ms of event-data (25 light pulses).
		The mean flow direction is from left to right.}
	\label{fig:3d_tracks}
\end{figure}

Fig.~\ref{fig:3d_tracks} shows two realizations of recovered near-wall tracks at $U_\infty = \qty{5.2}{\m\per\s}$ ($\mathrm{Re}_\tau = 563$). The particle positions are color-coded with the local wall stress magnitude  $\mid \vec{\tau}_w \mid$. 
While the tracks in Fig.~\ref{fig:3d_tracks}(a) indicate a low-shear condition and even some flow reversal, the flow topology is completely different only 8\,ms later (Fig.~\ref{fig:3d_tracks}b) when it is dominated by a high shear rate aligned with the mean flow direction. The shear rate partially exceeds the mean value by a factor of two.
Animations of the near-wall particle motion at different playback speeds are provided as part of the supplementary material, see Sec.~\ref{sec:supplementary}.

\begin{figure}[htb]
	\small
	\centerline{
		\includegraphics[width=\textwidth]{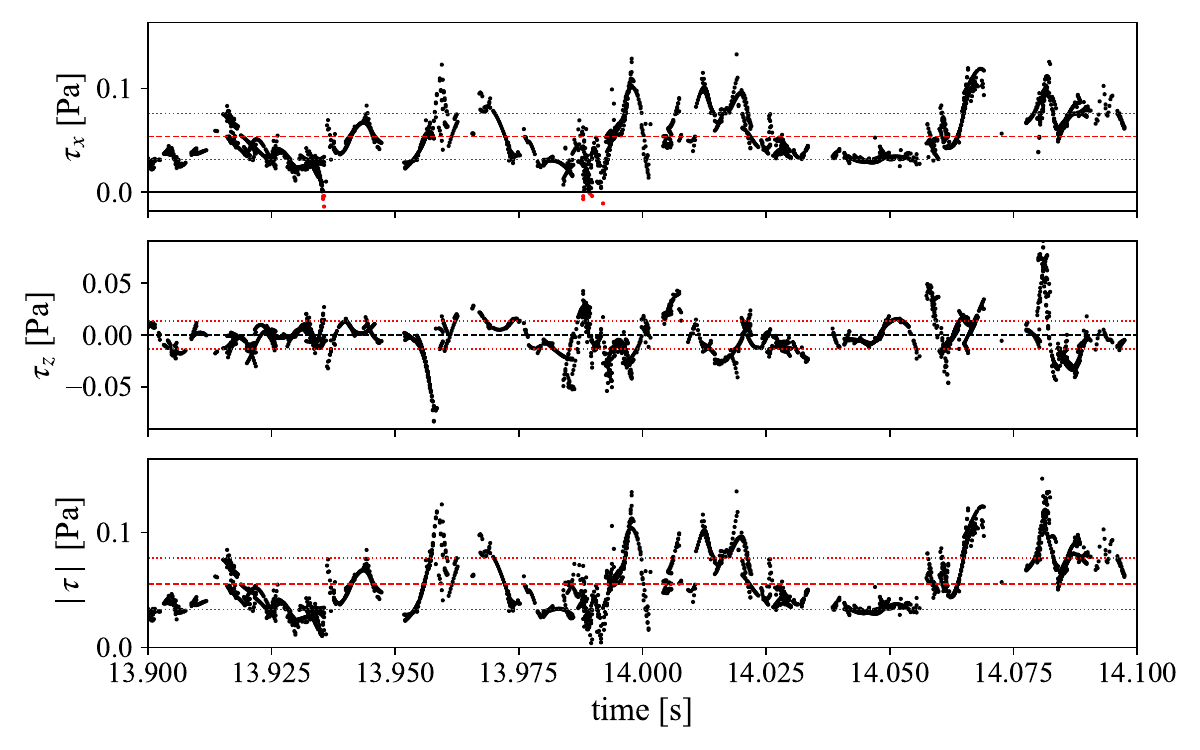}}
	\caption{\qty{0.2}{\s} sample of \ac{WSS}  data at $\mathrm{Re}_\tau = 563$ ($U_\infty = \qty{5.2}{\m\per\s}$) sampled in an area of $1\times 1 \,\unit{\square\mm}$, streamwise (top), spanwise (middle) and magnitude (bottom).
 Dashed lines indicates mean of quantity, dotted lines $\pm 1$ standard deviation (\ac{RMS}).
 }
	\label{fig:wss_vs_time}
\end{figure}

A short record of \qty{0.2}{\s} duration shows the evolution of \ac{WSS} estimates within a small ($1\times \qty{1}{\square\mm}$) area in Fig.~\ref{fig:wss_vs_time}.
The streamwise component $\tau_{x}$ reaches negative values (reverse flow, marked red in plot) at $t \approx \qty{13.93}{s}$ and $t \approx \qty{13.98}{s}$, whereas the spanwise component $\tau_{z}$ exhibits several extreme events in excess of 3-4 times the \ac{RMS}.
Of importance is the fact, that the randomness of the \ac{LPT} data does not provide continuous time records for a given point. This prevents the ad-hoc calculation of space-time characteristics such as frequency spectra or temporal correlations. 
To allow this, the random data would first need to be subjected to data assimilation or interpolation schemes.

\subsection{Mean velocity profile and statistics}
\begin{figure}[htb]  
   \includegraphics[width=\textwidth]{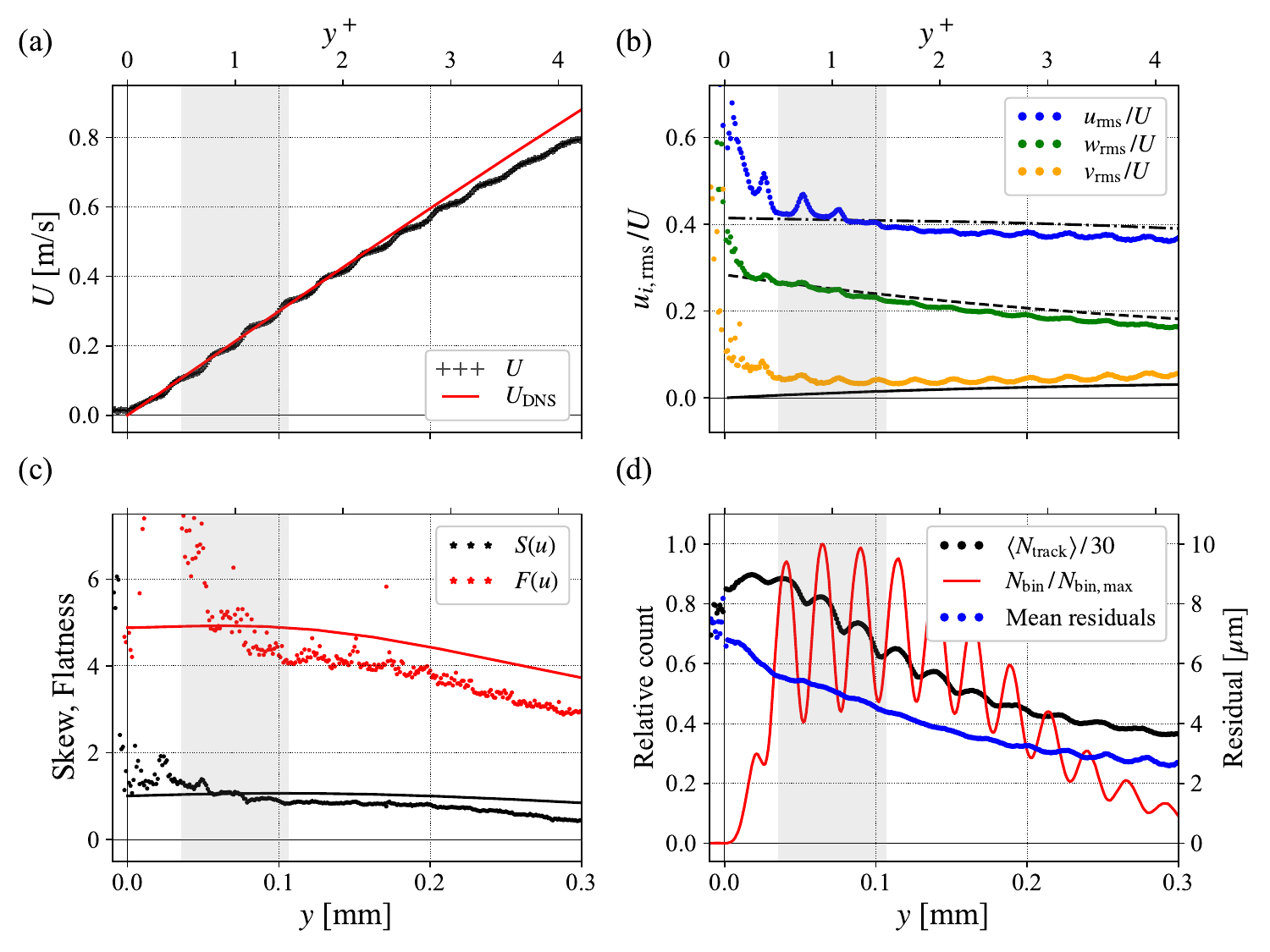}
    \caption{Bin-averaging results using bins of $\Delta y = 1\,\unit{\um}$: (a) near-wall velocity profile (+) and DNS prediction (red line),
    gray area indicates sampling domain for \ac{WSS} estimation;
    (b) \ac{RMS} of velocity components for streamwise (blue), spanwise (green) and wall-normal (gray) components;
	(c) skewness $S(u)$ and flatness $F(u)$ estimates of streamwise velocity in comparison to values predicted by DNS (solid lines, from \citealp{Schlatter:2010});
    (d) relative sample count $N_\mathrm{bin}$ within the bins (red line) and averaged track length normalized by maximum possible ($N_{\mathrm{max}} = 30$).
 }
    \label{fig:velocity-profile}
\end{figure}

Profiles of mean particle velocity and associated higher moments are compiled by bin-averaging across the \ac{FOV} at different discrete wall distances $y_i$ of sample height $\Delta y$.
Fig.~\ref{fig:velocity-profile} presents profiles obtained with a bin height of $\Delta y = \qty{1}{\um}$.
The mean streamwise profile (Fig.~\ref{fig:velocity-profile}a) is in good agreement with the DNS up to a wall distance of $y \approx 200\,\unit{\um}$ beyond which it begins to deviate. 
The deviation is believed to be sourced in the under-representation of faster particle tracks in the statistics: at a velocity of $U = \qty{1}{\m\per\s}$ the particles move \qty{20}{\px} between laser shots, such that faster moving particles are less likely to be tracked reliably.
For the present data set, useful velocities are available up to a wall distance of $y^+ \approx 4$ (\qty{300}{\um}).
For the estimation of \ac{WSS} the sampling volume is restricted to one viscous unit with the range $[0.5 < y^+ < 1.5]$.

The slope of the mean profile in a range $[20\,\unit{\um} \leq y \leq 150\,\unit{\um}]$ is a used for the estimation of the mean velocity gradient at the wall, ${\partial u}/{\partial y}\rvert_{y=0}$, which in turn is required in (Eq.~\ref{eq:shearstress}) to estimate the mean \ac{WSS}, $\langle \tau_w\rangle$, alongside with the estimation of the viscous scaling $l^* = \nu/u_\tau$.

The \ac{RMS} of all three components of the particle velocity are plotted in Fig.~\ref{fig:velocity-profile}(b) and are in reasonable agreement with DNS following the trend but slightly underestimate the DNS reference. 
The \ac{RMS} of the wall-normal component $v_\mathrm{rms}$ is at a nearly constant level throughout, indicating noise. 
With the near-wall flow essentially restricted to be only wall-parallel, track validation can rely on limiting the variance and magnitude of the wall-normal component (see also Fig.~\ref{fig:DNS_nearwall_rms}f).

Finally, Fig.~\ref{fig:velocity-profile}(c) provides the third and fourth order moments, that is, skewness $S(u)$ and flatness $F(u)$ of the streamwise velocity. 
While also underestimating the DNS predictions they begin to strongly deviate with increased proximity to the wall, indicating an increased amount of erroneous data near the wall, the net effect of which is averaged out in both the mean as well as the \ac{RMS} fluctuations. 

In addition to the profiles of mean and higher moments, Fig.~\ref{fig:velocity-profile}(d) provides the relative sample count for each of the \qty{1}{\um} bins (black line). 
This value modulates at a spatial frequency of $\Lambda = 24.0\pm \qty{0.5}{\um}$ and can be explained by an intensity modulation within the laser light sheet that is reflected by the glass surface while grazing it at a shallow angle. 
Within the darker regions, the probability of event generation is reduced which results in a local reduction of track data rate.
The spacing of the interference fringes, $\Lambda$, is related to the incidence angle $\theta$ by 
\begin{equation}
    \label{eq:interfreence}
    \Lambda = \dfrac{\lambda}{2 \sin(\theta)}    
\end{equation}
where $\lambda$ is the wavelength of the laser light (i.e. \qty{532}{\nm}).
The estimated angle of $\theta = 0.635\unit{\degree}$ matches the \qty{5}{\mm} entry height of the laser beam \qty{500}{\mm} away at the side of the tunnel.
The interference pattern introduces modulation in the velocity profiles and associated higher moments that become more pronounced with closer proximity to the wall.
Ideally this interference should have been prevented altogether, such as by placing a non-reflective coating just outside of the immediate field of view.
In light of this, the fluctuations $u_{i,\mathrm{rms}}$ have their highest deviations from the predicted profiles at the minimum sample count which suggests that the most reliable values are located at the maxima of the sample counts.
It should be noted that these modulations would not have been detected without an accurate plane adjustment as part of the camera calibration (cf. Fig.~\ref{fig:cam_views}b). 

\subsection{Wall shear stress distribution and statistics}
\label{sec:wss_stats}
\begin{figure}[htb]
    \includegraphics[width=0.95\columnwidth]{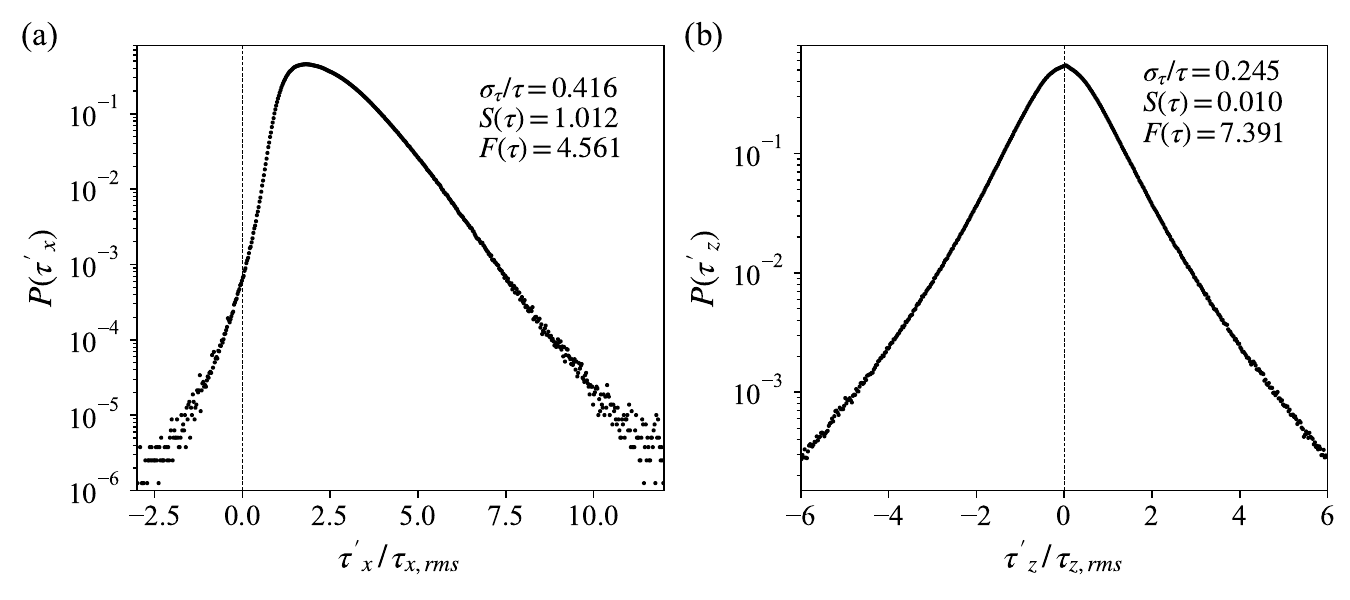}
    \caption{PDFs of streamwise (a) and spanwise (b) wall shear stress components normalized by the \ac{RMS} of the respective values compiled from $N_s = \num{30d6}$ correlated samples obtained from an event-record of 60\,s duration at $U_\infty = 5.2\,\unit{\m\per\s}$  
    ($\textrm{Re}_\tau = 563$).
    Velocity data is sampled in the range $[0.5 < y^+ < 1.5]$ using particle tracks with a minimal length of 7.}
    \label{fig:wallshear_pdfs5mps}
\end{figure}

\begin{figure}[htb]
    \includegraphics[width=0.95\columnwidth]{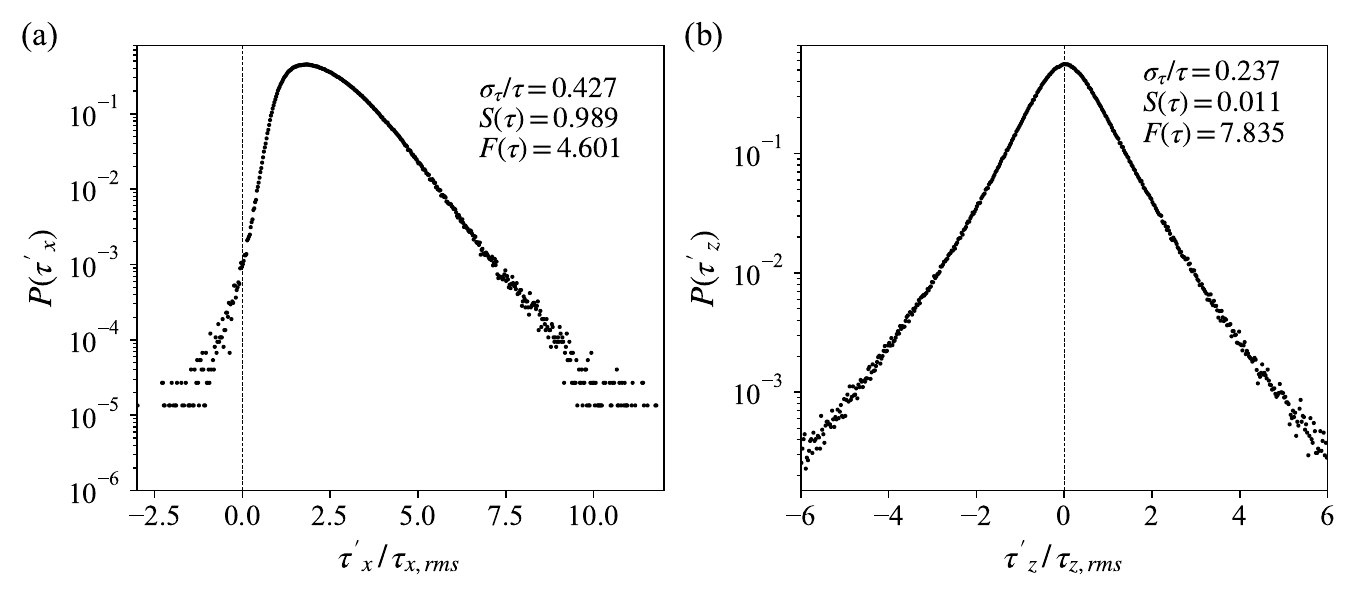}
    \caption{Same as Fig.~\ref{fig:wallshear_pdfs5mps} obtained at $U_\infty = 7.5\,\unit{\m\per\s}$ ($\textrm{Re}_\tau = 754$) using 10\,s of data with $N_s = \num{4.2d6}$.}
    \label{fig:wallshear_pdfs7mps}
\end{figure}

Following Eq.~\ref{eq:shearstress} the unsteady \acf{WSS} estimates are directly calculated using the particle's current velocity $u_i$ and distance from the wall $\Delta y_i$.
Probability distributions of both components of the \ac{WSS} vector are given in Fig.~\ref{fig:wallshear_pdfs5mps} for $U_\infty = 5.2\,\unit{\m\per\s}$ and Fig.~\ref{fig:wallshear_pdfs7mps} for $U_\infty = 7.5\,\unit{\m\per\s}$.
The distributions closely match those found in literature for DNS data (e.g. Fig.5 in \citealp{DiazDaniel:2017}) and experiments (e.g. Fig. 7 in \citealp{LiuSchroeder:2019}).
At $U_\infty = 5.2\,\unit{\m\per\s}$ the data set is based on a 60\,s record and sampled in a wall distance of $[0.5 < y^+ < 1.5]$ ($y = 35 \sim 105\,\unit{\um}$) for the estimation of the \ac{WSS} according to Eq.~\ref{eq:shearstress}. 
For 60\,s of processed event data the statistics represent a total of 8000 boundary layer turn-over times of $\delta_{99} / U_e = 5.2\,\textrm{ms}$ ($\mathrm{Re}_\tau = 563$, $\delta_{99} = 39\,\unit{\mm}$). 

The skewness $S_{\tau_x}$ and flatness $F_{\tau_x}$ of the \ac{WSS} components, summarized in Table~\ref{tab:wss_stats}, are in good agreement with data obtained at similar Reynolds numbers from DNS and experiments alike (see e.g. Table I in \citealp{DiazDaniel:2017}). 

\begin{table}[bt]
    \caption{Statistics of the \ac{WSS} for different data sets sampled at a wall distance $[0.5 < y^+  < 1.5]$. Estimated values for $\tau^+_{i,\mathrm{rms}}$ according to Eq.~\ref{eq:tau_corr}. 
    }
    \label{tab:wss_stats}
    \centering
    \begin{tabular}{l ccc|ccc|c}
    \toprule
    $\mathbf{U}_\infty$ $[\unit{m\per\s}]$ & \multicolumn{3}{c|}{\textbf{5.2}} & 
\multicolumn{3}{c|}{\textbf{7.5}} & \textbf{10.0} \\
    $\mathbf{Re}_\tau$  & \multicolumn{3}{c|}{\textbf{563}} & \multicolumn{3}{c|}{\textbf{754}} & \textbf{935} \\[3pt]
      Data set  & 5-3 & 5-1 & 5-4   & 7-1 & 7-2 & 7-3    & 10-1 \\
    \midrule
    $\tau^+_{x,\mathrm{rms}}$ (est.)   &
         & 0.412 &  &  & 0.417 &  & 0.421  \\
    $\tau^+_{x,\mathrm{rms}}$  &
        0.413 & 0.416 & 0.415   & 0.417 & 0.427 & 0.428   & 0.439 \\ 
    diff. & +\qty{0.2}{\percent} & +\qty{1.2}{\percent} & +\qty{0.7}{\percent} & $\pm\qty{0}{\percent}$ & +\qty{2.3}{\percent} & +\qty{2.6}{\percent} & +\qty{4.3}{\percent} \\[4pt]
    $\tau^+_{z,\mathrm{rms}}$ (est.)   &
         & 0.278 &  &  & 0.283 &  & 0.287 \\ 
    $\tau^+_{z,\mathrm{rms}}$   &
        0.243 & 0.244 & 0.240   & 0.235 & 0.237 & 0.237   & 0.229 \\ 
    diff. & -\qty{12.6}{\percent} & -\qty{12.2}{\percent} & -\qty{13.7}{\percent} & -\qty{17.0}{\percent} & -\qty{16.2}{\percent} & -\qty{16.2}{\percent} & -\qty{20.2}{\percent} \\[4pt]
    $S(\tau_x)$  & 0.978 & 1.012 &  1.027 & 0.989 & 0.997  & 0.992  & 0.942 \\
    $S(\tau_z)$  & 0.028 & 0.010 & -0.023 & 0.007 & -0.005 & 0.040  & -0.001 \\[4pt]
    $F(\tau_x)$  & 4.41  & 4.56  & 4.66   & 4.45 & 4.60 & 4.56     & 4.45 \\
    $F(\tau_z)$  & 6.76  & 7.39  & 7.77   & 7.43 & 7.83 & 8.19     & 7.20 \\
    $N_{s}$ 
    & \num{11e6}  & \num{32e6} & \num{26e6}  
    & \num{3.7e6} & \num{3.7e6} & \num{4.2e6} & \num{3.8e6} \\
    $N_{s,\mathit{eff}}$ 
    & \num{0.3e6} & \num{1e6} & \num{0.9e6} 
    & \num{0.2e6} & \num{0.2e6} & \num{0.3e6} & \num{0.4e6} \\
      \bottomrule
    \end{tabular}
\end{table}

The correlations for the \ac{RMS} of the \ac{WSS}, $\tau^+_{x_i,\mathit{rms}}$, proposed by \cite{OrluSchlatter:2011} have a Reynolds number dependency:
\begin{equation}
\label{eq:tau_corr}
	\tau^+_{x_i,\mathrm{rms}} = \dfrac{\tau_{x_i,\mathrm{rms}}}{\tau_w} = C_{0,i}+ 0.018 \ln \textrm{Re}_\tau
\end{equation}
with $C_{0,x} = 0.298$ and $C_{0,z} = 0.164$. 
At $\mathrm{Re}_\tau = 563$ this 
respectively predicts $\tau^+_{x,\mathit{rms}} = 0.412$ and $\tau^+_{z,\mathit{rms}} = 0.278$. 

\begin{SCfigure}[1][htb]
    \centering
    \includegraphics[width=0.6\columnwidth]{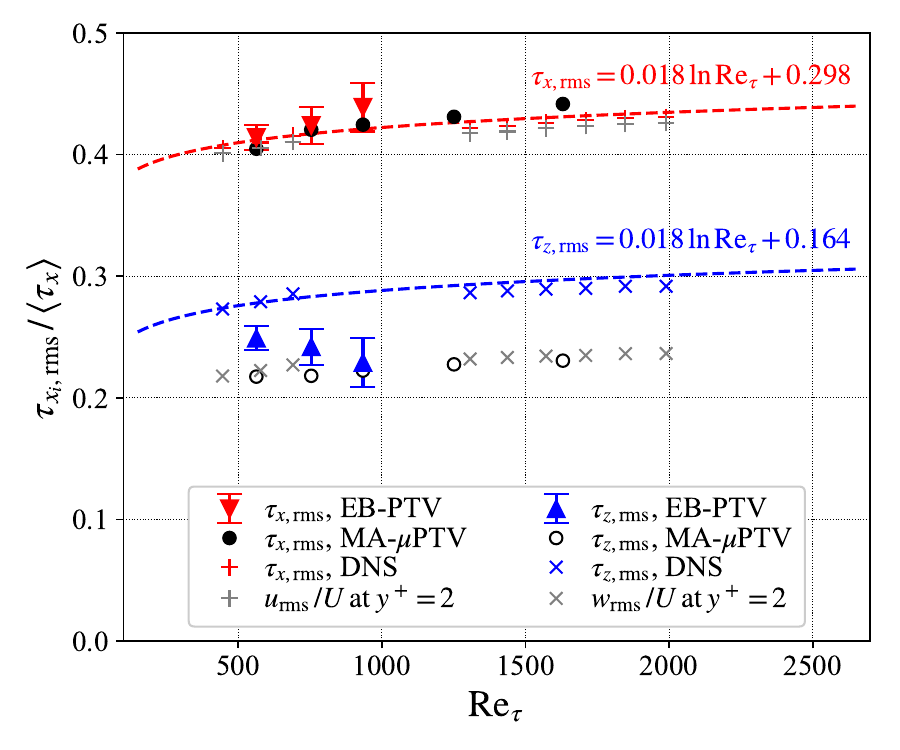}
    \caption{\ac{RMS} fluctuations of \ac{WSS} of streamwise (\textcolor{red}{ $\blacktriangledown$}) and spanwise (\textcolor{blue}{$\blacktriangle$}) components of the \ac{WSS}, determined from particle tracks in the viscous sub-layer for  $[0.5 < y^+ < 1.5]$.
    Round markers ($\bullet,\circ$) represent data obtained with MA-$\upmu$PTV \citep{Klinner:2024}; TBL-DNS ($+$, $\times$) by \cite{Sillero:2013}. Dashed lines correspond to Eq.~\protect\ref{eq:tau_corr} with different offsets $C_i$.}
    \label{fig:wallshear_rms}
\end{SCfigure}

The \ac{WSS} fluctuation estimates presented in Fig.~\ref{fig:wallshear_rms} are compiled from a variety of recordings at different seeding concentrations and different sampling intervals.
Whereas the streamwise \ac{WSS} fluctuations $\tau^+_{x,\mathit{rms}}$ are slightly overestimated, but within error bounds, the spanwise fluctuations $\tau^+_{z,\mathit{rms}}$ are underestimated by more than \qty{10}{\percent} which has also been observed in comparable measurements using \ac{MAmuPTV} \citep{Klinner:2024}.
A plausible explanation for this underestimation is the different convergence of the velocity fluctuations toward their limiting values at the wall as mentioned in Sec.~\ref{sec:error_estimation} (see Fig.~\ref{fig:DNS_nearwall_rms}f).

\begin{figure}[htb]
	{ \small{(a)}\hspace{0.49\textwidth}{(b)}}\\
	\centerline{
	\includegraphics[width=0.49\columnwidth]{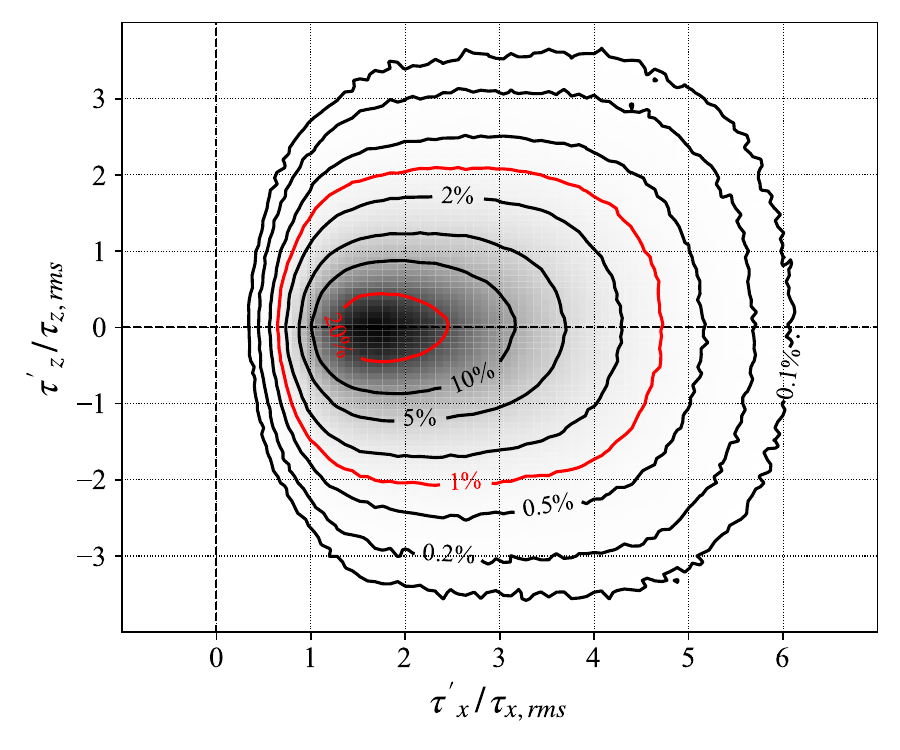}%
	\hfill
	\includegraphics[width=0.49\columnwidth]{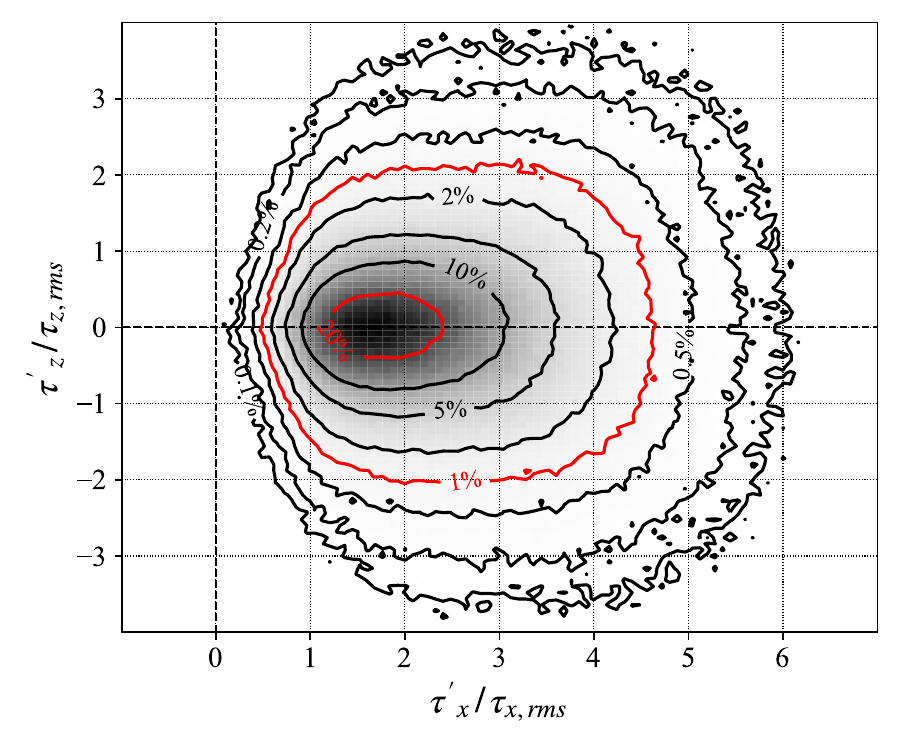}%
	}%
	\caption{Joint PDFs of the wall shear stress vector normalized by the \ac{RMS} of the respective components obtained at $U_\infty = 5.2$\,m/s ($\textrm{Re}_\tau = 563$, {(a)}) and $U_\infty = 7.5\,\unit{\m\per\s}$, ($\textrm{Re}_\tau = 754$, {(b)}). Contour levels represent probabilities of 0.1\%, 0.2\%, 0.5\%, 1\% (red), 2\%, 5\%, 10\%, 20\% (red). }
	\label{fig:wallshear_pdf2D}
\end{figure}

Joint probability distributions of the \ac{WSS} are plotted in Fig.~\ref{fig:wallshear_pdf2D} for two Reynolds numbers using a sample size of up to $N_s=30\times 10^6$ (at $\textrm{Re}_\tau = 563$).
These distributions agree very well with the results obtained with \ac{MAmuPTV} at the same Reynolds numbers (see \citealp{Klinner:2024}). Similar data has also been acquired by \cite{Sheng:2008} for turbulent channel water flow at a friction Reynolds number of $\textrm{Re}_\tau = 1400$.

The contours in Fig.~\ref{fig:wallshear_pdf2D}(b) show a small bulge near $\tau_w = 0$ which is believed to be caused by artefacts arising by the colinear arrangement of the three cameras along the streamwise direction. 
Due to this linear camera arrangement the epipolars between all three cameras are parallel. Therefore, multiple particles moving in streamwise direction have a higher likelihood of overlapping along the field of view and result in mismatching (ghost particles). 
This effect increases as the seeding density is increased. 
More sophisticated \ac{LPT} schemes, such as a modified 3d \ac{STB} \citep{Schanz:2016}, should be able to handle this deficiency and could potentially recover more tracks from the raw event-data. Adding a fourth camera would also reduce the likelihood of particle mismatch.

\subsection{WSS rate of change}
Visualizations of the particle motion within the viscous sublayer exhibit strong spanwise activity that give an impression that the spanwise unsteadiness is more pronounced and of higher amplitude than the streamwise fluctuations (see e.g. video supplement, Sec.~\ref{sec:supplementary}).  
To address this, we extracted the rate of change of the \ac{WSS} from the particle tracking data as part of the B-spline track fitting step.
Following the procedure of determining the unsteady \ac{WSS} vector from 
the near-wall velocity data as per Eq.~\ref{eq:shearstress}, the 
particle acceleration vector $\mathbf{a_p} = [a_{p,x}, a_{p,y}, a_{p,z}]$ can be retrieved from the tracking data and is related to the rate of change of the \ac{WSS} vector:
\begin{equation}
\label{eq:shearstresschange}
	\dfrac{\partial}{\partial t}\left(\vec{\tau}_w\right) 
    = \dfrac{\partial}{\partial t}\left(\mu \left.\dfrac{\partial \mathbf{u}}{\partial y}\right\rvert_{y=0} \right) 
    = \mu \left.\dfrac{\partial \mathbf{a}}{\partial y}\right\rvert_{y=0} 
	\approx \mu\, \dfrac{\mathbf{a_p}(\Delta y)}{\Delta y}.    
\end{equation}
In the following the symbol $\hat{a}$ refers to this quantity and in effect is the particle acceleration $a_{p,i}$ divided by the wall distance:
\begin{equation}
    \hat{a}_i = \frac{\partial \tau_i}{\partial t} = \lim_{y \rightarrow 0} \frac{a_{p,i}}{y}
\end{equation}
Similar to the wall-normal velocity component $v$, the wall-normal acceleration $a_y$ also vanishes due to the no-slip boundary condition at the wall.

\begin{figure}[htb]
\small
    (a)\hspace{0.3\columnwidth}(b)\hspace{0.3\columnwidth}(c)\\
    \includegraphics[width=\columnwidth]{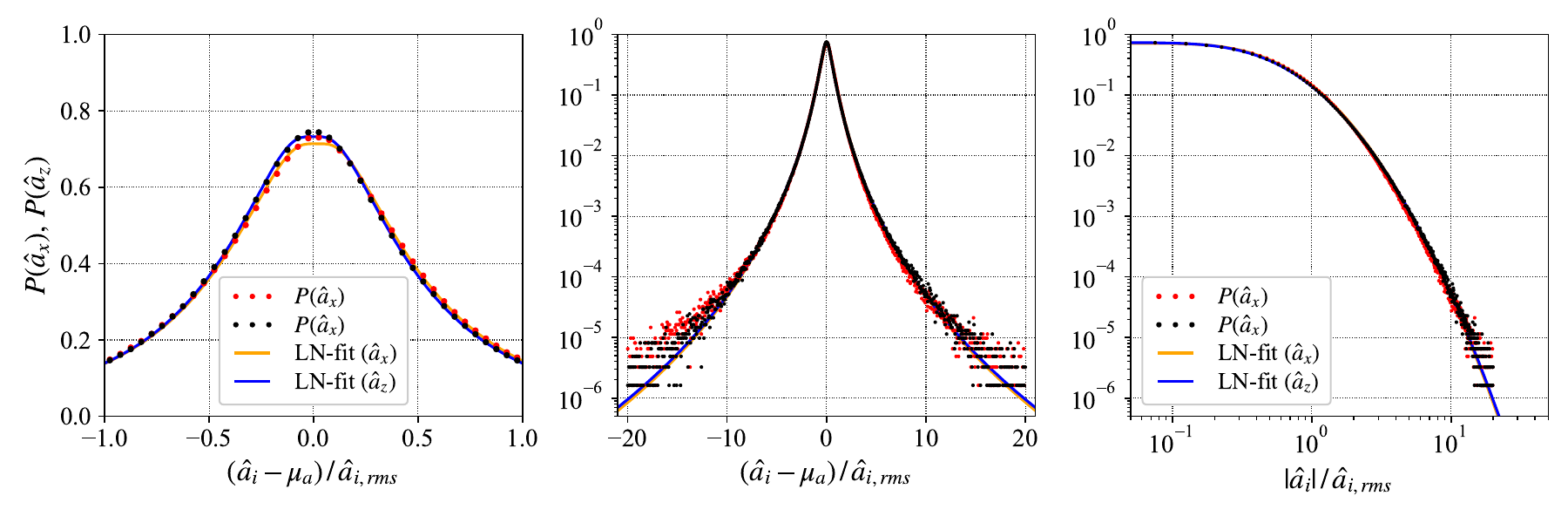}
    \caption{PDFs of rate of change of the \ac{WSS} components normalized by the \ac{RMS} of the respective values compiled from $N_s \approx 30\times 10^6$ samples at $U_\infty = 5.2\,\unit{\m\per\s}$ ($\textrm{Re}_\tau = 563$).
    Data is sampled in the range $[0.5 < y^+ < 1.5]$ using minimal track length of $N_\mathrm{track} = 7$. (a): detail near peak in linear scaling (b): logarithmic scaling,  (c) double-log scaling.
    Solid lines are stretched exponential fit according to Eq.~\ref{eq:expfit}.}
    \label{fig:acceleration_pdfs}
\end{figure}

\begin{figure}[htb]
	{ \small{(a)}\hspace{0.49\textwidth}{(b)}}\\
	\centerline{
	\includegraphics[width=0.49\columnwidth]{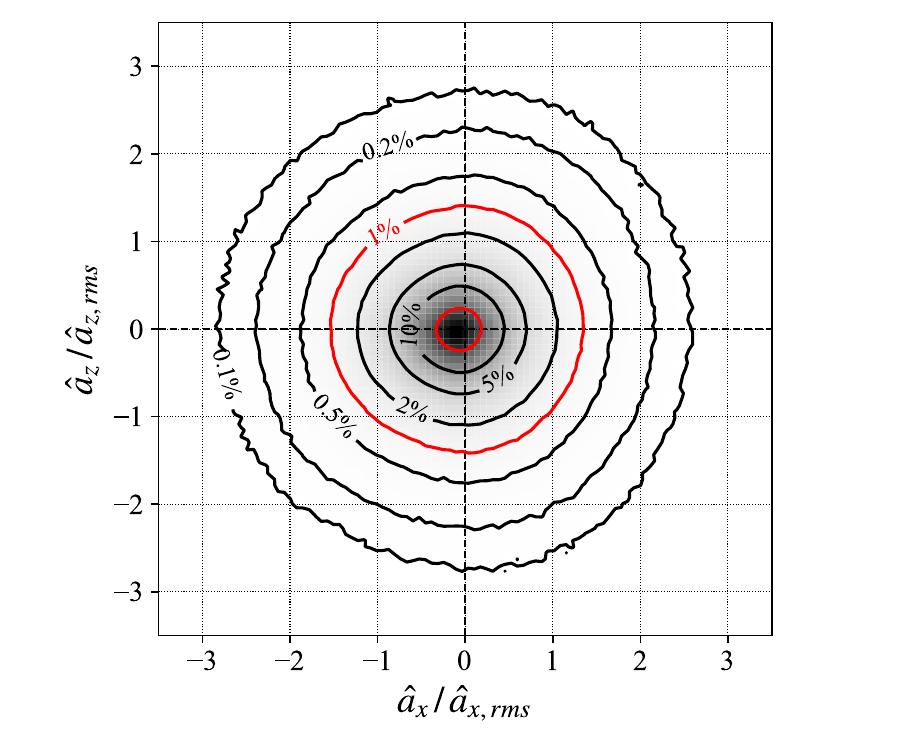}%
	\includegraphics[width=0.49\columnwidth]{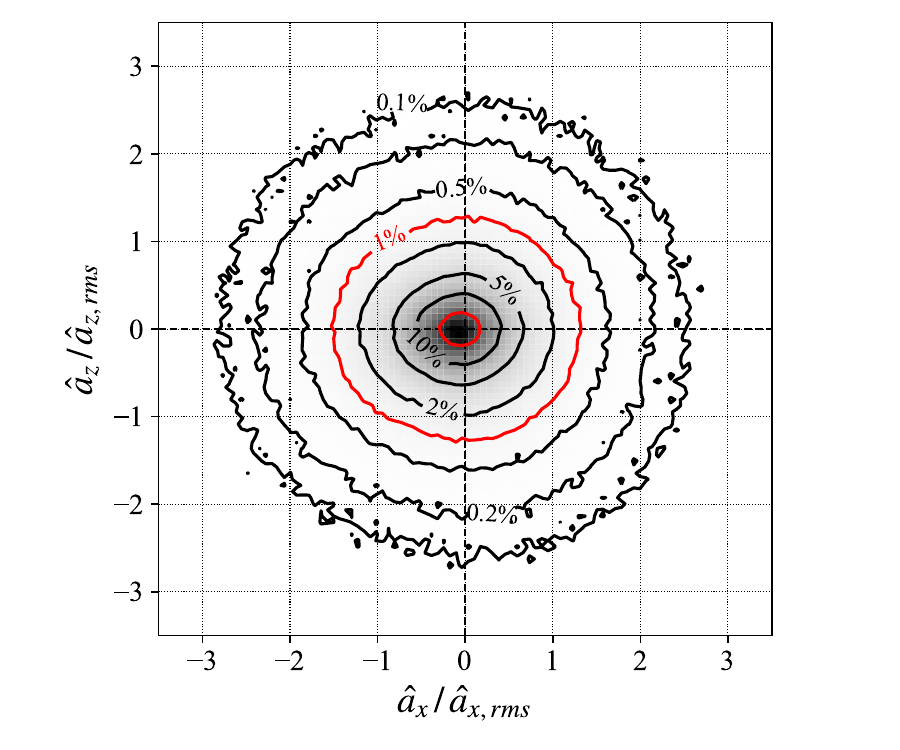}%
	}%
	\caption{Joint PDFs of the wall shear stress rate of change normalized by the \ac{RMS} of the respective components obtained at $U_\infty = 5.2\,\unit{\m\per\s}$ ($\textrm{Re}_\tau = 563$, {(a)}) and $U_\infty = 7.5\,\unit{\m\per\s}$ ($\textrm{Re}_\tau = 754$, {(b)}). Contour levels represent probabilities of 0.1\%, 0.2\%, 0.5\%, 1\% (red), 2\%, 5\%, 10\%, 20\% (red).}
	\label{fig:acceleration_pdf2d}
\end{figure}

Fig.~\ref{fig:acceleration_pdfs} presents \acp{PDF} of the rate of change of the \ac{WSS}, $\hat{a}$, for both streamwise and spanwise directions with a joint \ac{PDF} of the data shown in Fig.~\ref{fig:acceleration_pdf2d}. 
Noteworthy is the near perfect symmetry of the distributions with a slight shift of the streamwise component toward negative values (deceleration). 
This may be explained by the deceleration of the fluid within the boundary layer in the presence of a slightly positive pressure gradient. 
The \ac{RMS} of the \ac{WSS} rate of change $\hat{a}$ is also essentially equivalent for both components, deviating by less than \qty{5}{\percent} (cf. Table~\ref{tab:accel}).

\begin{table}[htb]
    \caption{Statistics of the \ac{WSS} rate of change $\hat{a}_i = \partial\tau_i\,/\,\partial t$ for different data sets sampled at a wall distance $[0.5 < y^+ < 1.5]$. Values for ``LN-fit" are obtained from nonlinear least-squares fit of Eq.~\ref{eq:expfit}.
    }
    \label{tab:accel}
    \centering
    \begin{tabular}{ll ccc|ccc|c}
    \toprule
    $\mathbf{U}_\infty$, $[\unit{m\per\s}]$ & & \multicolumn{3}{c|}{\textbf{5.2}} & 
\multicolumn{3}{c|}{\textbf{7.5}} & \textbf{10.0} \\
    $\mathbf{Re}_\tau$ & & \multicolumn{3}{c|}{\textbf{563}} & \multicolumn{3}{c|}{\textbf{754}} & \textbf{935} \\[3pt]
      Data set & & 5-3 & 5-1 & 5-4 & 7-1 & 7-2 & 7-3 & 10.1 \\
    \midrule
    $\langle \hat{a}_x \rangle$, $[\unit{Pa\per\s}]$ & &
        -0.849 & -0.866 & -0.900   & -3.055 & -2.901 & -2.979 & -6.227 \\
    $\langle \hat{a}_z \rangle$, $[\unit{Pa\per\s}]$ & &
        -0.018 & -0.006 & -0.005   &  0.001 & -0.032 & 0.023 & -0.006 \\[2pt]
    $\hat{a}_{x,\mathrm{rms}}$, $[\unit{Pa\per\s}]$ & & 
      11.42 & 10.92 & 11.32    & 34.47 & 33.82 & 34.02   & 75.44 \\
    $\hat{a}_{x,\mathrm{rms}}$, $[\unit{Pa\per\s}]$ & & 
      11.17 & 10.83 & 10.96    & 34.18 & 34.15 & 34.47   & 72.93 \\
    $\frac{\hat{a}_{z,\mathrm{rms}}}{\hat{a}_{x,\mathrm{rms}}}$ &  &
      0.978 & 0.976 & 0.968    & 0.991 & 0.976 & 1.01   & 0.967 \\[5pt]
    $S(\hat{a}_x)$ &  &
      -0.628 & -0.883 & -0.862 & 0.600 & -0.110 & -0.275 & -0.184 \\
    $S(\hat{a}_z)$ &  &
      -0.024 & 0.009  & -0.056 &  0.060 & 0.275 & -0.000 & 0.219 \\[4pt]
    $F(\hat{a}_x)$ &  &
    49.4 & 58.5 & 58.2   & 65.3 & 28.3 & 47.7    & 27.5 \\
    $F(\hat{a}_z)$ &  &
    32.8 & 29.9 & 51.5   & 35.5 & 75.1 & 31.8    & 74.1 \\
    $F(\hat{a}_x)$ (LN-fit) &  & 32.7 & 33.5 & 33.9  & 34.9 & 34.6 & 34.7 & 35.4 \\
    $F(\hat{a}_z)$ (LN-fit) &  & 34.7 & 35.9 & 37.1  & 41.9 & 41.4 & 42.1 & 46.8 \\[4pt]
    $s_x$ (LN-fit)  &  & 0.770 & 0.772 & 0.782   & 0.807 & 0.802 & 0.802 & 0.829 \\
    $s_z$ (LN-fit)  &  & 0.782 & 0.791 & 0.811   & 0.848 & 0.849 & 0.853 & 0.921 \\
    $m_x$ (LN-fit)  &  & 0.977 & 0.959 & 0.956   & 0.954 & 0.959 & 0.959 & 0.964 \\
    $m_z$ (LN-fit)  &  & 0.936 & 0.913 & 0.904   & 0.848 & 0.848 & 0.836 & 0.819 \\
      \bottomrule
    \end{tabular}
\end{table}

Distributions of the near-wall particle acceleration and its relation to the \ac{WSS} rate of change are rarely reported in the literature and mostly discussed in the context of inertial particle transport using \ac{DNS}.
The PDFs shown in Fig.~\ref{fig:acceleration_pdfs} are strongly non-Gaussian and exhibit strongly pronounced tails with a high flatness $F(\hat{a}) = 30\sim 50$, which is indicative of high intermittency.
This was already observed by e.g. \cite{YeoKimLee:2010} for \ac{TCF} with $\mathrm{Re}_\tau = 180 \sim 600$. 
The distributions in Fig.~\ref{fig:acceleration_pdfs} have a strong resemblance to acceleration data obtained from both \ac{DNS} as well as experiments (by \ac{LPT}) for inertial particles in \ac{HIT} \citep{Voth:2002,Mordant:2004,Schroeder:Lisbon:2022}.
In this context, a stretched exponential function, also known as log-normal superstatistics \citep{Beck:2004}, is generally used to describe the shape of the probability distribution with \cite{Stelzenmuller:2017} proposing the following expression:
\begin{equation}
   p(\hat{a}_i) = \frac{e^{s_i^2/2}}{4m_i}
   \left[ 
   1 - \mathrm{erf}\left( 
   \frac{\ln{\frac{\mid \hat{a}_i \mid}{m_i} + s_i^2}}{\sqrt{2}\, s_i}
   \right)
   \right].
   \label{eq:expfit}
\end{equation}
According to \citep{Stelzenmuller:2017} the parametric variable $s_i$ defines the shape of the distribution whereas $m_i$ is related to the variance of $a_i$.
For the present \ac{WSS} rate of change data, nonlinear least-squares fitting yields $s = 0.785\pm 0.014$ and $m = 0.941\pm 0.026$ and an estimated flatness in the range of $30\sim 40$ (c.f. Table~\ref{tab:accel}).
Although the investigated near-wall flow is very anisotropic by nature, the striking similarity of the PDFs $P(\hat{a})$ to those of \ac{HIT} could indicate a certain universality as already pointed out by \citep{Stelzenmuller:2017}.
Whereas previous studies have noted an increased asymmetric PDFs of the acceleration components with increased proximity to the wall (e.g. \citealp{Stelzenmuller:2017}, \citealp{Zamansky:2011}, \citealp{YeoKimLee:2010}), there is little information on the limiting case of particle acceleration very close to the wall ($y^+ < 2$ in present experiments), in particular, the rate of change of the \ac{WSS}, $\hat{a}$, determined from the particle acceleration (see Eq.~\ref{eq:shearstresschange}).
Our findings suggest a fully symmetric (isotropic) behavior of the wall shear stress rate of change.


\subsection{Two-point correlations of \ac{WSS}}
\label{sec:2pt_corr}
Two-point correlations provide a measure of similarity between the data $u_1$ at a given point $\mathbf{x}_0$ in space $\mathbf{x}$ (or time $t$) with respect to the data point $u_2$ in its neighborhood.
Here it is calculated for the two wall shear stress components $\tau_i$\,=\,$\tau_i$\,($x,y,z,t$) using the discrete version of the cross-correlation coefficient:

\begin{eqnarray}
R_{\tau_i \tau_j}(\mathbf x, \mathbf x_0)
    &=& \frac{\int \tau_i'(\mathbf x, t) \, \tau_j' (\mathbf x_0, t) \, dt}
{\sqrt{\int \tau_i'(\mathbf x, t)\, \tau_i' (\mathbf x, t)\, dt} \cdot \sqrt{\int u_j'(\mathbf x_0, t)\, \tau_j' (\mathbf x_0, t) \, dt}} \\
    &=& \frac{\langle\, \tau_i'(\mathbf x, t) \cdot \tau_j' (\mathbf x_0, t)\, \rangle}{ \langle \tau_i'(\mathbf x, t)^2\rangle^{0.5} \cdot \langle \tau_j'(\mathbf x_0, t)^2\rangle^{0.5} }
    \label{eq:corr2pt}
\end{eqnarray}
\noindent%
where the two quantities in the denominator are the square roots of the sample variances (i.e \ac{RMS} of $\tau_k$) while $y$ is held constant (i.e near wall plane, $y^+ = 1$).
With the present \ac{WSS} data being ungridded, the calculation of Eq.~\ref{eq:corr2pt} requires a bin-averaging approach.
Furthermore, it is assumed that the flow statistics are constant across the field of view such that each sampled value at position $\mathbf x_0$ is assumed to be located at the origin $(x=0, z=0)$.
The distance to other points in the sample defines the location of the bin for incremental accumulation of the correlation statistics.
As a result the effective size of the correlation map is larger than the data domain, with decreasing bin entries toward the edges.

For the present data set, a square bin size of $\Delta x \times \Delta z = 250 \times  250\,\unit{\square\um}$ ($3.6\,x^+ \times 3.6\,z^+$) was chosen.
As in the previous data processing the data is sampled from a volume of one viscous height $[0.5 < y^+ < 1.5]$ resulting in a sample of $N_s \approx 30\times 10^6$.

\begin{figure}[htb]
	\centerline{
	\includegraphics[width=0.49\columnwidth]{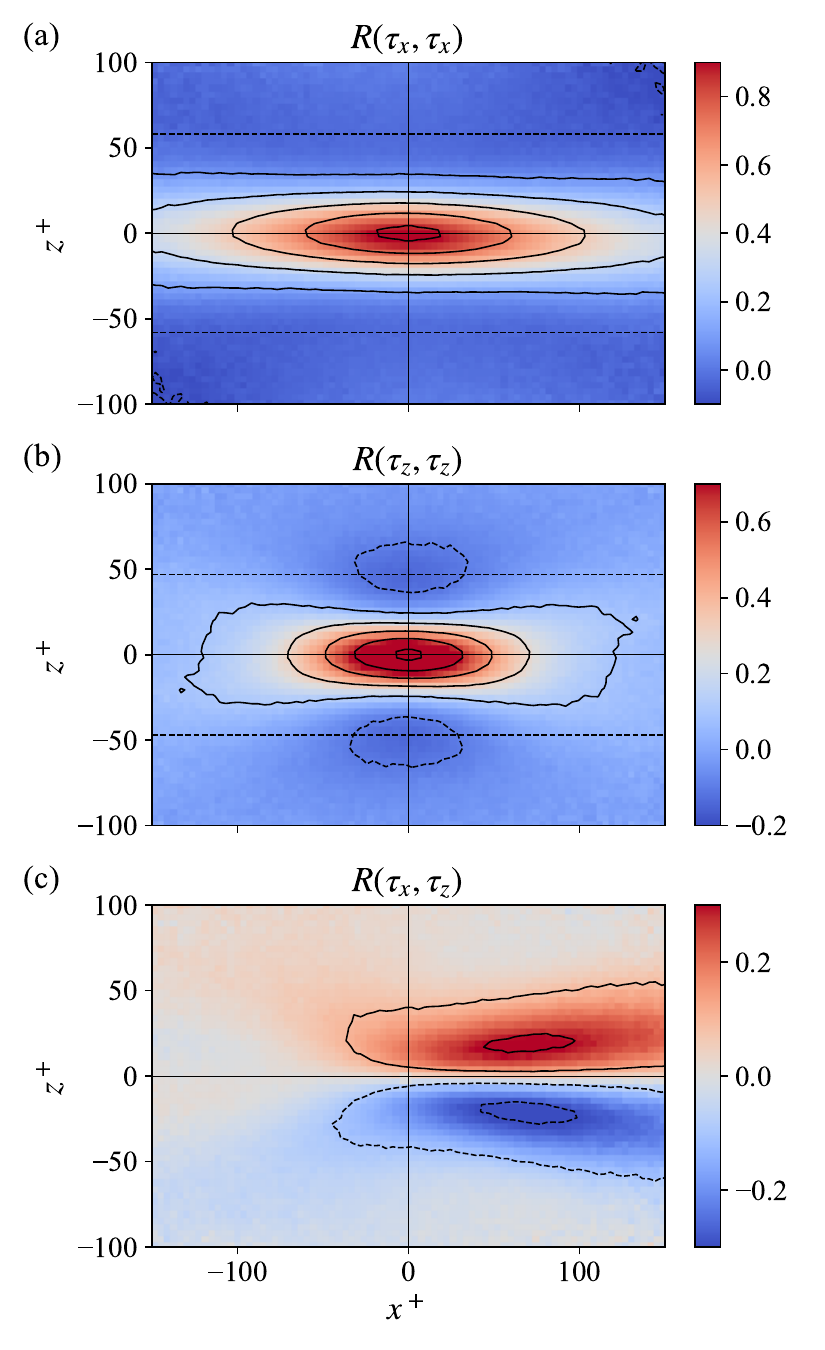}%
	\includegraphics[width=0.49\columnwidth]{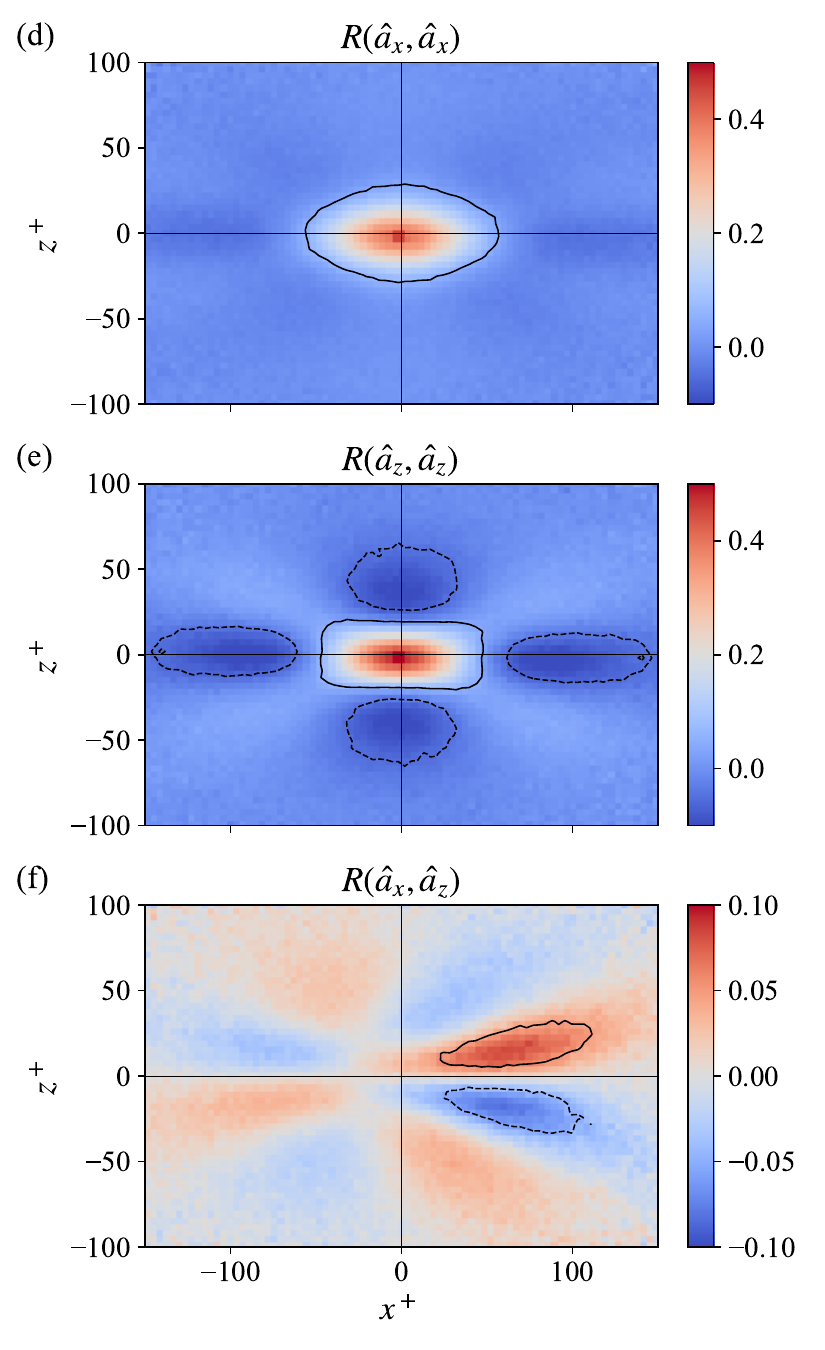}%
	}%
	\caption{Two-point correlation maps of the \ac{WSS} (a)-(c) and its rate of change $\hat{\mathbf{a}}$ (d)-(f) obtained at $U_\infty = 5.2\,\unit{\m\per\s}$  ($\textrm{Re}_\tau = 563$) using tracking results from $[0.5 < y^+ < 1.5]$ and sample size of $N_s \approx 30\times 10^6$. Contour lines at $-0.1$, $+0.1$, $+0.3$ $\ldots$ (left column), $\pm0.05$ (right column). Horizontal dashed lines indicate position of minima along $x = 0$. }
	\label{fig:2ptcorr}
\end{figure}

The two-point correlation maps provided in Fig.~\ref{fig:2ptcorr}(a-c) agree with DNS-based results by \cite{JeonMoin:1999} for $\mathrm{Re}_\tau = 180$ with deviations most likely related to the difference in Reynolds number (see Figs. 7a,b and 8a in their paper). 
Recent 3d-\ac{STB} data by \cite{Schroeder:Lisbon:2024} obtained from the same wind tunnel facility yielded very similar two-point correlation maps, albeit the sampling plane being located at $y^+ = 5$.

The elongated contours of $R_{\tau_x \tau_x}$ are related to the streamwise near-wall streaks -- wall-parallel, counter-rotating vortical structures aligned in streamwise direction. 
Along $x=0$ the minima are respectively located at $\pm58\,z^+$ for $R_{\tau_x \tau_x}$ and $\pm47\,z^+$ for $R_{\tau_z \tau_z}$ which corresponds to the mean spanwise spacing of about $100\sim 120$ viscous units reported in the literature (\citealp{SmithMetzler:1983} and others).
The correlation map for $R_{\tau_x \tau_z}$ shows a double-peak feature inclined at $\approx\pm 5\unit{\degree}$ that relates the streamwise \ac{WSS} $\tau_x$ to an off-axis maximum spanwise $\tau_z$ about $70\sim 80\,x^+$ further downstream. 
This topology is likely to be related the $\approx\pm 6\unit{\degree}$ features observed in space-time correlations by \cite{Lagraa:WSS:2004}, although this needs further investigation.

Compared to the correlation maps of \ac{WSS}, the spatial signature of the \ac{WSS} rate of change, $\hat{a}$, shown in Fig.~\ref{fig:2ptcorr}(d-f), is much more compact with lobe-like negative correlation features. 
Very similar topology has been reported by \cite{Schroeder:Lisbon:2024} (see Fig. 11 in their publication).
The fluid dynamical processes associated with the correlation topologies have yet to be investigated in further detail and literature on this particular aspect is not known to the authors. 


\section{Discussion}

In the course of the experiments event data was collected at different seeding densities to assess its influence while keeping all other parameters constant.
The track detection rate given in Table~\ref{tbl:event_data} indicate that 
there is an optimum event data rate of about 8-\qty{12e6}{Ev/s}. A further increase of the event rate (= higher seeding) actually results in a reduction of the track detection rate. 
With increased particle image density the likelihood of particle ambiguity and false track initialization also increases. 
At the highest seeding level with more than \qty{20e6}{Ev/s} (data set 5-2) the valid track validation rate drops to less than \qty{5}{\percent}, which is why this data set was omitted in the data analysis. 
Here an approach that first reconstructs the 3d particle positions followed by 3d track building, rather than tracking in 2d space for each camera view, is likely to provide better results.

Along with the increase in event-rate, the saturation of the sensor readout causes increased latency in the time stamping such that pulses are no longer clearly separated; derived pseudo-frames will contain particle images from more than one pulse which cannot be separated in time. This also impacts the particle tracking performance.

Even at optimal particle image density and event data rate, the \ac{WSS} determined from the tracking results showed a consistent underestimation of the spanwise \ac{WSS} fluctuation $\tau^+_{z,\mathit{rms}}$. 
This was also found in related measurements using highly accurate micro particle tracking techniques by \cite{Kumar:2021} and \cite{Klinner:2024}.
Here DNS is particularly helpful in explaining the underestimation:
Fig.~\ref{fig:DNS_nearwall_rms}(d) provides profiles of the velocity fluctuations for all 3 velocity components. Focusing in on the near-wall region ($y^+ < 8$) in Fig.~\ref{fig:DNS_nearwall_rms}(e) they are characterized by different rates of change, with $u^+_\mathrm{rms}$ to strongest, followed by spanwise $w^+_\mathrm{rms}$ ($\approx 40\unit{\percent}$ at $y^+ = 5$) and wall-normal $v^+_\mathrm{rms}$ ($\approx 10\unit{\percent}$ at $y^+ = 5$).
However, when these quantities are normalized with the mean streamwise velocity $U(y)$ as shown in Fig.~\ref{fig:DNS_nearwall_rms}f, they exhibit a completely different behavior: while the quantity $u_\mathrm{rms}/U$ shows gradual decrease, its spanwise counterpart $w_\mathrm{rms}/U$ rapidly decreases with increasing wall distances, whereas the wall-normal quantity gradually increases from zero at the wall. 
The limiting values of the former two quantities, $u_\mathrm{rms}/U$ and $w_\mathrm{rms}/U$, at the wall ($y=0$), in fact, coincide with the \ac{WSS} fluctuations and represent the DNS-based estimates in Fig.~\ref{fig:wallshear_rms}.
In the context of velocimetry-based \ac{WSS} estimation, the velocity must be sampled at a finite distance $\Delta y$ from the wall. Close to the wall, both the velocity and wall-distance approach zero and relative errors rapidly increase as explained in Sec.~\ref{sec:error_estimation}.
Since the quantity $u_\mathrm{rms}/U$ has a weaker decay compared to $w_\mathrm{rms}/U$, the latter will always be underestimated to a much higher degree.
This is illustrated in Fig.~\ref{fig:wallshear_rms} by sampling the DNS data at a finite wall distance of $y^+ = 2$ as indicated by the gray symbols. This sampling domain is comparable to that chosen for the \ac{WSS} estimation in the present work and leads to a comparable underestimation of the spanwise \ac{WSS} fluctuation $\tau^+_{z,\mathit{rms}}$.
In principle the under-estimation can be corrected by computing the velocity variances at different wall-distance intervals and extrapolating the trend toward the wall.
The velocity fluctuations plotted in Fig.~\ref{fig:velocity-profile}(b) closely follow the DNS predictions and justify the extrapolation approach.

At the highest bulk velocity of $U_\infty = 10\,\unit{\m\per\s}$ the particle track yield was insufficient for reliable \ac{WSS} estimation, in part, due to the nearly doubled mean particle displacement (compared to $U_\infty = 5.2\,\unit{\m\per\s}$), but also, because of the proportional reduction of the viscous scale from $\nu/u_\tau = 69\,\unit{\um}$ to $\nu/u_\tau = 37\,\unit{\um}$.
To a certain extent, a proportionally higher laser pulsing frequency could improve the measurement. However, the bandwidth limitation of the \ac{EBV} camera hardware imposes a limit of about \qty{10}{\kHz}, in particular, at increased seeding levels.
Overall it was found that the data quality improves with reduced seeding density which is related to the improved particle matching using only 3 cameras. Adding a fourth camera in the setup would provide additional redundancy, stabilizing the 3d particle position reconstruction. 

In terms of \ac{FOV} and spatial resolution the herein introduced configuration has advantages over other \ac{WSS} measurement techniques reported in the literature.
Covering an area of $12\times 7.5\,\unit{\square\mm}$ ($170x^+ \times 110z^+$), the \ac{FOV} of the present implementation is considerably larger than that of the micro-pillar technique ($2.1 \times 2.1\,\unit{\square\mm}$ \citealp{LiuSchroeder:2019})
or \ac{DFRH} ($1 \times 1\,\unit{\square\mm}$, $20x^+ \times 20z^+$, \citealp{Kumar:2021}) and \ac{muDH} ($1.5 \times 1.5\,\unit{\square\mm}$, $88x^+ \times 88z^+$, \citealp{Sheng:2008}).
Similarly the depth-from-defocus approaches have a small \ac{FOV} on the order of $1 \times 1\,\unit{\square\mm}$ \citep{Fuchs:WSS:2023,Klinner:2024}.
The MEMS based \ac{WSS} ``imagers" by \cite{Kimura:MEMS:1999} provided a \ac{FOV} of $22 \times 7.5\,\unit{\square\mm}$, however, on a relatively coarse grid of sensors consisting of 3 rows spaced at $\Delta x = \qty{10}{\mm}$ with 25 sensors each spaced at $\Delta z = \qty{300}{\um}$.
In this regard, the present work offers both a high spatio-temporal resolution on a \ac{FOV} covering in excess of one mean wall-streak spacing.

\section*{Conclusion \& Outlook}
\label{Sec:Conclusions}
The material presented herein demonstrates the viability of event-based imaging velocimetry for accurate measurement of \ac{TBL} properties by means of Lagrangian particle tracking, providing near-wall velocity profiles and \ac{WSS} distributions along with derived quantities. 
The reduced data stream of \ac{EBV} permits continuous recording on the order of minutes (or longer) using off-the-shelf computer systems for data storage.
Uncertainties arising from the limited (1-bit) signal depth of the image data are accounted for by making use of the available temporal resolution of the raw data which is on the order of 5--\qty{10}{\kHz}.
Track reconstruction can be greatly improved using Wiener or Kalman filtering such as implemented by \cite{Borer:2017}. 

Even without processing, the raw event-data is well suited for the visualization of the near wall dynamics. 
While this is also possible with high-speed particle imaging approaches, the inherent binary nature of the raw imagery captured by event cameras immediately provides high contrast visualizations without additional effort (see e.g. event data animations provided in the supplementary material, Sec.~\ref{sec:supplementary}).
In the present application, rapid spanwise modulations imparted by the passage of flow structures in the outer layers of the \ac{TBL} are clearly visualized and suggest further spatio-temporal analysis of the dynamics to retrieve, for instance, the structure convection velocity.

The time-resolved data presented herein was acquired using hardware that is considerably cheaper in comparison to conventional high-speed \ac{PIV} components necessary to achieve similar results and but, at this point, are unable to stream images for extended periods. 
Beyond this, the higher sensitivity of the \ac{EBV} detectors reduce the power requirements of the laser used to illuminate the tracer particles.

The present measurements were performed at a laser pulsing frequency of \qty{5}{\kHz}. Although not discussed here, a small portion of data was also acquired at 10\,kHz and provided acceptable results in spite of a partial leakage (overflow) of some events into the following laser pulse period.
Given the same magnification and the frequency limit of about 10\,kHz for the utilized event camera hardware, the proposed technique should be applicable to \ac{TBL} flows with friction velocities approaching $u_\tau = \qty{1}{\m\per\s}$.

\backmatter
\bmhead{Acknowledgments}
We are very appreciative of the support by our colleagues of DLR Institute of Aerodynamics and Flow Technology during our measurement campaign in G\"{o}ttingen, in particular Prof. Markus Raffel and his team for the use of the wind tunnel facility
and Prof. Andreas Schr\"{o}der for providing the high-speed cameras for the high-speed profile-PIV reference measurements.

\section*{Declarations}

\begin{itemize}
%
\item \textbf{Ethics approval:}
Not applicable
\item \textbf{Competing interests:}
There are no competing interests to declare.
\item \textbf{Funding:} This work was made possible through DLR-internal funding.
\item \textbf{Availability of code, data and materials:} Sample event data can be obtained from the author upon request. 
\item \textbf{Authors' contributions:}
CW conceptualized and realized the event-based 3d tracking measurement technique. CW and JK jointly set up and performed the wind-tunnel measurements, and subsequently worked out the data processing strategies, event data processing codes and accompanying data analysis; CW wrote the original draft of the manuscript, which was reviewed by JK and department colleagues.
\end{itemize}

\begin{appendices}
\section{Supplementary material}
\label{sec:supplementary}
Animated sequences of the acquired event data and recovered near-wall particle tracks are provided as supplementary material.
\begin{itemize}
    \item File \textbf{Suppl1-Events-Vel5mps-ts5ms-0.1x.mp4} - visualization of events captured by central camera at $U_\infty = 5.2\,\unit{m\per\s}$ at $0.1\times$ actual speed; time-slice of 5\,\unit{\ms} per pseudo-image (25 pulses per image), event rate \qty{7.0e6}{Events\per\s}, positive events only.
    \item File \textbf{Suppl2-Events-Vel5mps-ts5ms-0.01x.mp4} - visualization of events captured by central camera at $U_\infty = 5.2\,\unit{m\per\s}$ at $0.01\times$ actual speed; time-slice of 5\,\unit{\ms} per pseudo-image (25 pulses per image), event rate \qty{7.0e6}{Events\per\s}, positive events only.
    \item File \textbf{Suppl3-WSS-magnitude-200ms.mp4} - visualization of near-wall particle tracks color coded with magnitude of the \acf{WSS}. Only 3 most recent time steps of tracks are color-coded, then fade from gray to black ($U_\infty = 5.2\,\unit{m\per\s}$, speed about $0.02\times$ actual speed).
\end{itemize}
\end{appendices}

\bibliography{EBIV_lit}

\end{document}